\documentclass[prb,floatfix,twocolumn,showpacs,a4paper,nobibnotes,nofootinbib]{revtex4}
\usepackage{graphics}

\begin{document}

\title{Kinetic energy driven stripe formation and pairing in repulsive electronic systems}
\author{Marco Bosch}
\affiliation{Institute Lorentz for Theoretical Physics, Leiden University\\
P.O.B. 9506, 2300 RA Leiden, The Netherlands}
\author{Zohar Nussinov}
\affiliation{Department of Physics, Washington University, St. Louis,
MO 63130, U.S.A.}

\date{\today}
\email{zohar@wuphys.wustl.edu}

\begin{abstract}
We study repulsive 
Hubbard and $t-J$ type systems on a square lattice (long believed to capture certain quintessential aspects of
the high temperature superconductors). These models (alongside the parent compounds of the high temperature superconductors) are antiferromagnetic in the absence of hole doping. As we illustrate, a unifying underlying principle for the dynamics of holes introduced by doping rationalizes the emergence of nonuniform electronic structures-- ``stripes'' and possible
pairing tendencies therein. Specifically, our analysis invokes the following (numerically verified) sublattice parity principle:
{\it a strong antiferromagnetic background forces injected
holes to hop in steps of two such that they always
remain on the same sublattice}.
When applied to a domain
wall in an antiferromagnet, 
this simple principle naturally gives rise to (bond centered) stripes. 
We demonstrate that the holes are self-consistently
 localized on stripes. Extending this picture, we then show that the holes
on a stripe favor the formation of 
pairs on neighboring rungs or sites. 
Throughout this work much emphasis is placed 
on the problem of a two leg ladder 
immersed in a staggered magnetic
field. Although we will focus on the square lattice,
our considerations may be extended to similar electronic structures appearing in other models on bipartite lattices
when these exhibit antiferromagnetic correlations with an underlying sublattice structure.
\end{abstract}

\pacs{71.g}
\maketitle

\section{Introduction}

The current work constitutes an updated version of ideas and results concerning the possibility of ``{\it kinetically driven confinement}''  in the cuprate superconductors that we introduced in [\onlinecite{us2002}]. Three decades have passed since the discovery of these high-\( T_{c} \)
superconductors. \cite{BM86} Although much has 
been learned since, there is still no satisfactory
explanation of what
causes the superconductivity. There is a widespread belief
that it should be possible to describe many electronic properties of
the square CuO\( _{2} \) lattices by the bare two-dimensional repulsive
Hubbard type model \cite{jh,h1} when endowed with additional longer range hopping terms and interactions,
\begin{equation}
\label{Hubbard model}
H=-t\sum _{\langle ij\rangle \sigma }(c_{i\sigma }^{\dag }c_{j\sigma
}+c_{i\sigma }^{\dag }c_{j\sigma })+U\sum _{i} n_{i\uparrow }n_{i\downarrow} + \cdots,
\end{equation}%
where \( c_{i\sigma }^{\dag } \) creates an electron on site \( i \)
having a spin component \( \sigma  \).
The number operators $n_{i \sigma} = c_{i \sigma}^{\dagger} c_{i \sigma}$.
In its original bare incarnation (sans the additional terms denoted by an ellipsis in Eqn. (\ref{Hubbard model})) \cite{jh}, 
the Hubbard model contains both the hopping of electrons between neighboring sites (\( t \)) and a repulsive on-site Coulomb energy penalty (\( U \)). 
The Hubbard model is one of the simplest possible models 
of interacting electrons. The main problem
is then finding the possible solutions
of this model. This however has proven not to be an easy task. The
Hilbert space of the model is too large for exact numerical solutions.
Away from half filling, Monte Carlo simulations have the minus sign problem. \( U \) is approximately
equal to \( 8t \), so we are neither in an extremely strong, nor in
a weakly interacting limit. Therefore, standard perturbative techniques may be of limited
use. Intense investigations employing high temperature series expansions, infinite dimensional dynamical mean field
results, renormalization group calculations, exact quantum Monte Carlo calculations at half-filling, 
and various approximate numerical approaches have employed to the study of this
model \cite{h1,study1,study2,study3,study4,study5,study6,study7,study8,study9,study10,study11,WS}.
In the limit  of large $U/t$, a perturbative expansion relates the Hubbard model to another extremely well studied system,
the so-called ``$t-J$ model'' \cite{tj'}, 
\begin{eqnarray}
H_{t-J} =
-t  \sum_{\langle i j \rangle, \sigma}   (c_{i, \sigma}^{\dagger} c_{j, \sigma}
+ H.c.) +  J \sum_{\langle i j \rangle}  \vec{S}_{i} \cdot
\vec{S}_{j}.
\label{tJ}
\end{eqnarray}
Here, $\vec{S}_{i} = \sum_{\sigma \sigma^{\prime}} c_{i
\sigma}^{\dagger} \vec{\sigma}_{\sigma, \sigma^{\prime}} c_{i,
\sigma^{\prime}}$
denotes the electron spin operator at site $i$ (the vector
$\vec{\sigma}$ represents the Pauli matrices). The number operator $n_{i} = n_{i \uparrow} + n_{i \downarrow}$. 
The $t-J$ model Hamiltonian of Eqn. (\ref{tJ}) is defined on a Hilbert space in which $n_{i}  \le 1$. To leading order in a perturbative expansion of the Hubbard model in $(t/U)$, the exchange $J$ in Eqn. (\ref{tJ}) is $J=\frac{4t^2}{U}$. Away from the regime of large $U/t$ (far from the antiferromagnetic phase), the $t-J$ model is quite distinct from the Hubbard model. 
Variants of the $t-J$ model (including the ``$t-J_{z}$'' model) are known \cite{gerardo} (and in some cases, e.g., \cite{I1} can be rigorously demonstrated) to exhibit intricate correlations and structures including paricular ``stripe orders'' that will form the focus of the current work. Many aspects of the $t-J$, Hubbard, and other related models have been investigated throughout the years \cite{sri,other1}.
Numerically, energy differences between contending low energy stripe and other states are often found to be small , e.g., \cite{corboz}. Disputes appear in the literature as to which of the suggested states are more accurate and of lower energy. For irrational doping fractions in related systems, the stripes may form quasi-crystalline structures with gapless excitations \cite{eranzohar}. The large number of
contending low energy states and associated conflgurational entropy suggest that glassy dynamics might be possible \cite{sw,zis}. 
 In order to obtain results using simple illuminating analysis, that do not rely on elaborate calculations, and may rather universally rationalize the appearance of low energy stripe like structures, we will invoke simplifications.
The main assumption that we will rely on (and establish in the appendix) is that, when lightly doped, these systems 
are endowed with sublattice constraints on the kinematics. The nearly degenerate states of pristine Hubbard or $t-J$ models with no additional terms may be of academic interest. A broader approach, such as the one that we follow here, centers on a prominent effect captured by these systems. In the current work, we will, for concreteness, perform simple calculations applied to Hubbard models. Our considerations may apply to a broader class of systems that have a bipartite N\'eel antiferromagnetic (or other) background that imposes restrictions on hole dynamics. 

Stripes were first detected in cuprates\cite{tranquada} 
in the famous neutron scattering
experiment of Tranquada and coworkers. 
Earlier theoretical approaches already predicted stripes before their
experimental discovery \cite{jan,steve}. Various aspects have been investigated by numerous works, e.g., \cite{mya,seamus}.
Notably, similar phenomena and viable physical underpinning may also appear in further unconventional (pnictide and other) superconductors different from the cuprates, e.g., \cite{seamus}. 
A plot of the incommensuration as a function of doping (the Yamada plot), which is a straight line from
0 to 0.125 doping with a slope of \( 1/4 \) \cite{yamada}, indicates that stripes may be quarter filled.
DMRG calculations by
White and Scalapino \cite{WS} have further indicated that the ground-state
for stripes in the $t$-$J$ model is approximately quarter-filled.  
In the intervening years, many additional facets having been discovered.
Related issues that have become the focus of intense investigations
include the specter nematic orders and the relation between stripes, nematic order,
and superconductivity \cite{vojta,Achkar}. Aside from the initial appearance of
the current work \cite{us2002} on stripes within Hubbard type models, there have been many works that attempted to relate
stripes and related charge, spin, and pair density wave and nematic orders as well as superconducting tendencies \cite{h1,sco1,sco2,sco3,sco4,Gull,sco5,sco6,sco7,sco8,sco9,sco10,sco11,sco12,sco13,sco14,sco15,sco16,sco17,sco18,sco19,sco20,sco21,sco22,sco23,sco24,comin,bosch} to these models. In an offshoot of this paper, we detail a possible reason why stripes are quarter-filled \cite{us2}. 
As alluded to above, the principal goal of the current work is to rationalize how the square lattice Hubbard model of Eqn. (\ref{Hubbard model})
is amenable to stripe formation and pairing. As we will expand on, our guiding principle is that of a sub lattice constraint on hole motion;
this constraint is expected to arise in bipartite systems that display N\'eel like order in their half-filled state (the state in which there is one
electron per lattice site). We will see that this will kinetically favor the formation on stripes structures in which the holes aggregate together
in intricate the domain well stripe structures so as to allow nearest neighbor hopping along the stripe. As our point of departure relies on
the constraints imposed by an antiferromagnetic background, our considerations cannot be extended to certain pyrochlore lattice and other 
Hubbard type systems that do not exhibit conventional antiferromagnetic N\'eel orders but instead exhibit spin liquid or other ground states \cite{nn2015,pyron}. In this paper, we will assume
from the outset that bond-centered
stripes (stripes with the geometry of two 
leg ladders) form antiphase domain
walls in a surrounding N\'eel type antiferromagnetic background 
and demonstrate how self consistent localized eigenstates and 
pairing follow from this assumption.

\section{outline}

The remainder of this work is organized as follows:
In Section (\ref{principle}), we 
introduce the sublattice principle 
which will form the backbone of
our analysis. In order to streamline the quintessential physics, we will be
brief in exposing this principle. 
For further details regarding the underpinnings of this principle, the interested
reader may peruse Appendix A.
Invoking the sublattice principle, we
will show in Section (\ref{confine})
how exponentially localized wavefunctions
will be found when we assume, self consistently, that stripes
form antiphase domain walls. We will find
that the transverse stripe scale is of 
the order of the lattice constant. 
What drives stripe formation
in our picture are not confining magnetic string potentials,
but a rather novel kinetic effect which we term 
{\em{dynamical confinement}}.

Once we establish, self consistently,
that bond centered stripes form
domain walls in the surrounding 
antiferromagnet, we will move
in small steps towards examining
further microscopics.
As we will explain in Section (\ref{LaDDeR}), we 
will consider a bond centered
stripe engulfed by a surrounding antiferromagnetic
region as a two leg ladder immersed in an external
staggered magnetic field.
We will then consider 
the problem of a single electron on a
staggered empty two leg ladder (Section (\ref{empty-one}))
which will
facilitate the analysis of
a single hole on an otherwise
full staggered two leg ladder (Section (\ref{full-one})). Both problems
will lead to similar results.

This will be followed by a similar analysis
for a two electron system on an 
empty staggered ladder and the inverted 
problem of two holes on 
a full staggered ladder, in
Sections (\ref{2-empty})
and (\ref{2-full})
respectively. The surprising conclusion is that in this case there
is an essential difference between electrons and holes.
Numerically, we find
that for holes, pair states
are slightly favored 
over single hole states
although the correlations
are very faint. 

Having shown how narrow
bond centered stripes with
bound pair states emerge, 
in Section (\ref{DMRRRG})
we fuse all of 
our finding together 
to reconstruct earlier
suggested stripe patterns found by DMRG, mean-field, 
and many other methods. 

In Section (\ref{added}),
we will examine the effects of
longer range Coulomb effects and 
additional longer range kinetic terms to show
how much of our self consistent
analysis can easily
be fortified by
the addition of 
such terms.

\section{The Sublattice Parity Principle}
\label{principle}

As has long been appreciated \cite{h1,study1,study2,study3,study4,study5,study6,study7,study8,study9,study10,study11,WS,vojta,Achkar,sco1,sco2,sco3,sco4,Gull,sco5,sco6,sco7,sco8,sco9,sco10,sco11,sco12,sco13,sco14,sco15,sco16,sco17,sco18,sco19,sco20,sco21,sco22,sco23,sco24,comin,bosch}, the problem posed by the two-dimensional Hubbard model of Eqn. (\ref{Hubbard model}) is very rich. To make progress in a cogent way, in this article, we will invoke the simplifying assumption that much of
the low energy physics of holes in a strong
antiferromagnetic background can be summed 
in a nutshell: 

\[ \fbox {Holes\, can\, only\, move\, in\, steps\, of\, two.}\]

This principle is equivalent to
the omission of spin flips, waves,
and string states.
This sublattice parity principle is, 
of course, a gross oversimplification- the physics
of hole motion in an antiferromagnetic
is a fascinating and rich topic. Nonetheless,
the low energy single hole dispersion curves (requiring 
careful extensive numerical work) coincide 
very well with those immediately following 
from this principle.
For a review of this often 
overlooked principle,
the reader is invited 
to read Appendix (\ref{sub_prin}). 
 
Stated in formal terms, the principle amounts to
$Z_{2}$ order, which has lately been 
much discussed in the 
context of its viable destruction \cite{zohar}.

Many beautiful works exist on the
subject of stripe formation and dynamics \cite{erica}.
Hole dynamics on the stripe has been primarily addressed
in terms of spinon-holon excitations, as in the
works of Tchernyshyov and Pryadko
\cite{pryadko}. 
In this article we do {\em not} assume
that the elementary excitations are spinons and holons.
Our approach is closer in spirit to the
very interesting work of Chernyshev, 
White, and Castro Neto \cite{strings}, which
aim to answer the same fundamental questions concerning stripes.
Their work examines the elementary excitations
within the framework of the $t$-$J_{z}$ model by 
examining retraceable paths of a hole out of the stripe
into the surrounding antiferromagnet.
We arrive at much of the same physics using the sublattice parity principle
as our only guide. Our derivations are much more pedestrian yet physically
transparent as compared to the detailed Green's function and 
numerical DMRG analysis carried out by these authors.

Many works place
much emphasis on the string
states created by hole
motion or on kinks \cite{wim}. 
In our low energy analysis, there are {\em no string
states}. For a 
discussion on how string
states can be avoided and 
for a demonstration that they 
do not dominate 
the low energy physics, 
the reader is referred to Appendix (\ref{sub_prin}).
Similarly, the role of a domain wall in an antiferromagnet
as an effective attractive potential 
for holes simply by the bad ferromagnetic
magnetic bonds that they remove, and
the careful interplay between
string states and transverse
and longitudinal hole kinetics
is not what we consider
here. A large body of literature
complements our simple 
asymptotic low energy analysis.

\section{A Bond centered stripe in the antiferromagnet}
\label{local}

\begin{figure}
{\centering \resizebox*{0.8\columnwidth}{!}{\includegraphics{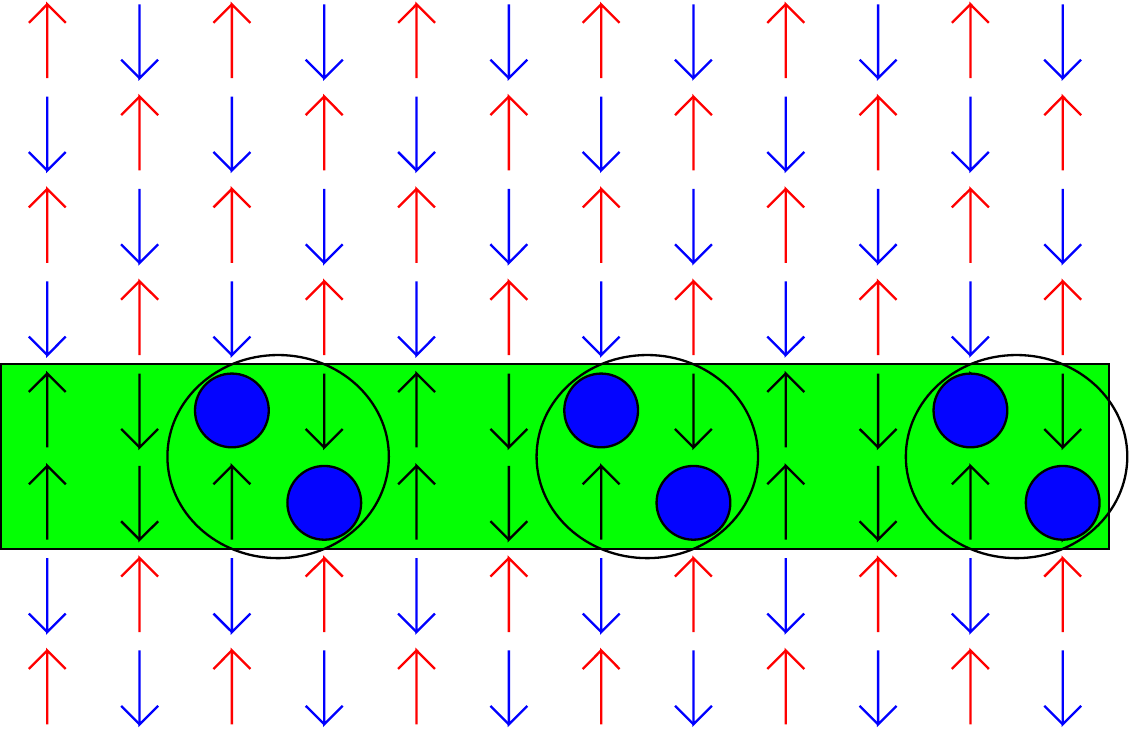}} \par}
\caption{Schematic representation of a quarter-filled bond-centered stripe
in an antiferromagnet. We assume there are no spin-flips in the antiferromagnet.
There is a \protect\( \pi \protect \)-phase shift in the staggered
magnetization over the stripe. The cartoon above is
not to be taken too literally. In reality, 
the pairs are smeared along the rungs.}
\label{figure: AF+bond-centered stripe with 3 pairs (the Model)}
\end{figure}

Fig. (\ref{figure: AF+bond-centered stripe with 3 pairs (the Model)})
shows the system which we will be looking at. We will examine the
anatomy 
of a bond-centered
stripe in an antiferromagnetic background. As foretold, we will 
neglect spin-flips and assume the background to be a perfect
N{\'e}el antiferromagnet. We will further assume that there is a phase-shift in
the staggered antiferromagnetic order parameter as we traverse the
stripe. This implies that there is a seam of ferromagnetic bonds between
spins on the two different legs of the stripe. The ferromagnetic
bonds will cost a lot of energy. However, introducing holes into the
stripe will reduce this strain. The ferromagnetic bonds 
will turn out to be essential to explain the stability 
of stripes against ``hole evaporation''.

\begin{figure}
{\centering \resizebox*{0.7\columnwidth}{!}{\includegraphics{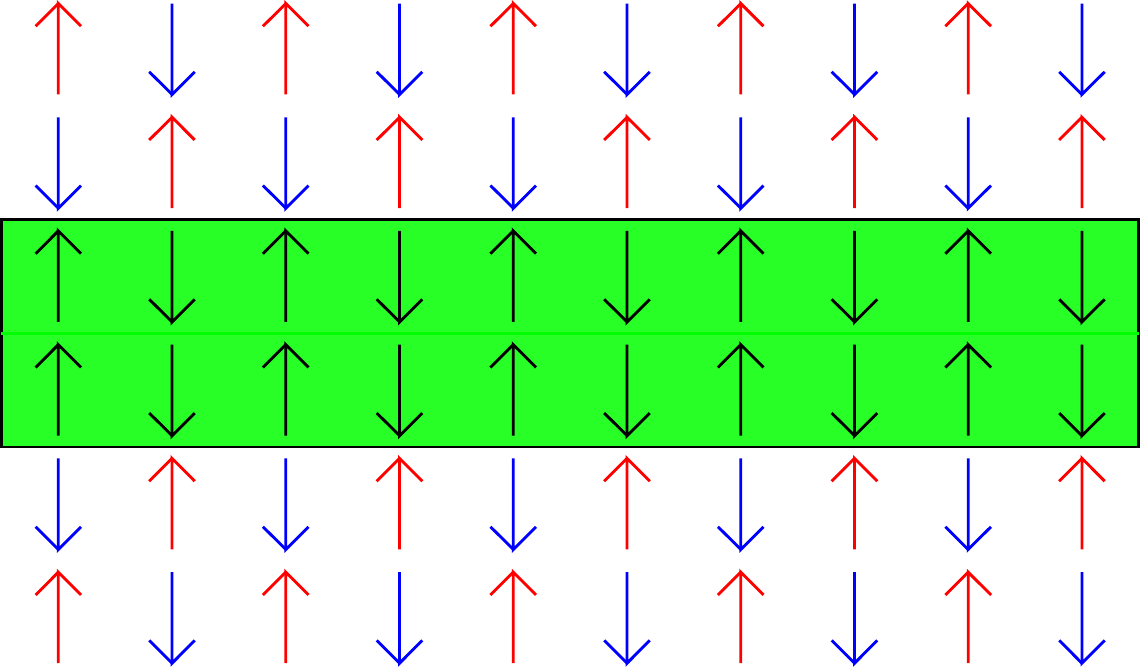}} \par}
\caption{We will make the assumption that the system starts out as antiferromagnetic
with a ``{\em domain wall}'' - a skyrmion. We will analyze 
what happens if we add holes to
this system.\label{figure: AF with a domain wall without holes}\label{figAFdomainWallNoHoles}}
\end{figure}

In order to understand why a system would want to form such a stripe, 
we make the theoretical assumption
that the system starts out with an antiphase boundary as shown in
Fig. \ref{figure: AF with a domain wall without holes}. Theoretically,
one can force such a system by imposing suitable boundary conditions.
If the system is \( L \) sites long, this will entail an
energy penalty of, approximately, \( 2LJ_{z} \). Introducing 
holes into this system will reduce the energy of this ferromagnetic seam 
and ameliorate matters.

Topologically, stripes are not literally domain walls in an 
antiferromagnet (whose spins $\vec{S}$
are continuous variables and {\em 
not} Ising
like) but rather topological excitations of the antiferromagnet
known as skyrmions. 
As noted by Wilczek and Zee \cite{wilzee} and later incorporated 
by others in rather novel theories, e.g. Wiegmann \cite{paul},
skyrmions in an antiferromagnet
are cylindrical domains separating N{\'e}el
states shifted by half a period. We note that topologically, 
stripes are identical
to skyrmions stretched out to 
form domain walls spanning the entire lattice. 
Berry phase effects in the antiferromagnet associated with
domain wall stripes can
give rise to exotic statistics similar to that of 
skyrmions in 2+1 dimensions- as suggested 
by extending the results of  
\cite{wilzee} to 
our domain walls. 
Subsequent the initial appearance of our work \cite{us2002},
further illuminating investigations modeled magnetic order and fluctuations
via an analysis of coupled two leg ladders \cite{konik}.

\begin{figure}
{\centering \resizebox*{1.03\columnwidth}{!}{\includegraphics{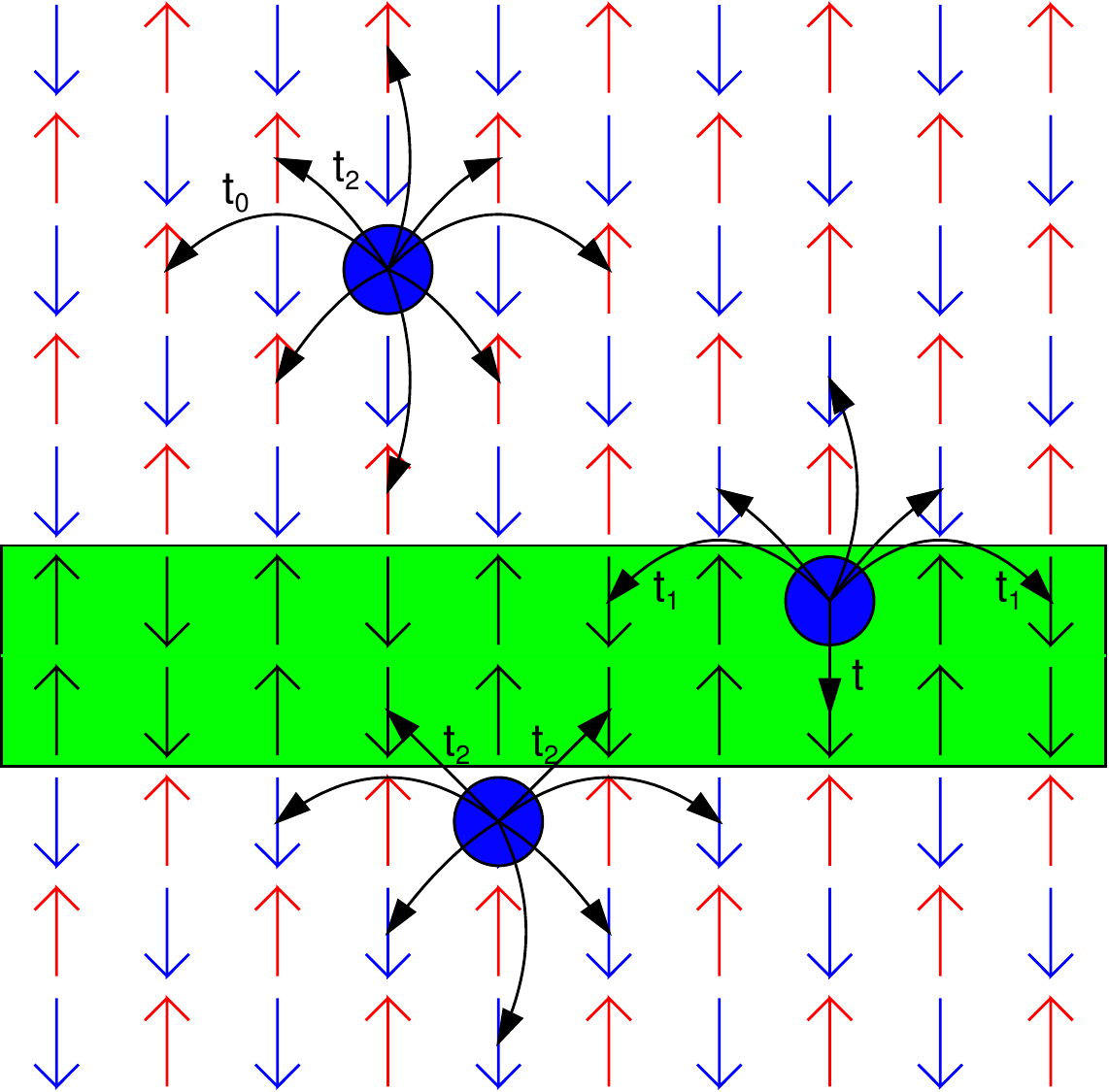}} \par}
\caption{A number of holes and an antiferromagnetic
{\em domain wall} immersed in the surrounding 
antiferromagnet. We will show that the holes will automatically end up
on the domain wall because of the single step up/down movement with 
an amplitude
$t$ which is possible within the domain wall
(the stripe), but
not in the surrounding antiferromagnet.\label{figure: Hole in AF + domain wall}}
\end{figure}

\section{Dynamical Confinement (on-stripe hypothesis)}
\label{confine}

We now examine what happens if we introduce a hole into the antiferromagnet
with a domain wall (see Fig. \ref{figure: Hole in AF + domain
wall}). Once again, our major assumption still is that this hole moves
in steps of two such that it will always stay on the same sublattice. 
We reiterate a note for the experts: 
our analysis has nothing to do with string states 
and kink (wiggle) motion beautifully analyzed
by many \cite{strings,pryadko}, and 
many of the works in 
\cite{beginn,a1,a2,a3,a4,a5,Louis,a6,a7,a8,a9,a10,a11,a12,a13,endd}
(our point of departure -the sublattice parity principle- 
is just a way to explicitly avoid high
energy string states from
the outset). Nor, does 
our bare analysis have anything to do with the role 
of the magnetic alleviation energies 
for a hole drifting into the stripe 
(whereby it removes bad 
ferromagnetic bonds)
which 
plays the role of an effective
attractive potential. 
Such effects will only enhance the bare findings 
that we report here.

With these remarks in mind, we note that if the hole 
is in the antiferromagnet, far away from 
the stripe, it has eight positions were it can move to 
(see Fig. \ref{figure: Hole in AF + domain wall}).
If it is sitting next to the boundary (just outside the stripe), it
loses one hop, because of the reversed sublattice structure. The remaining
seven hops are all the same as if the boundary was not there. 

Larger changes happen with a  
hole located within the stripe. For such a hole there are only six hops left. In Fig.
\ref{figure: Hole in AF + domain wall}, the hole on the stripe cannot move
downward left or downward right because of the change in
sublattice structure. So, naively, one might come to the conclusion
that a hole has more kinetic energy in the antiferromagnet then if
it is on or near the boundary/stripe. Naively, confinement of the hole in the
stripe would cost kinetic energy. However, this is not the case.

\begin{figure}
{\centering \resizebox*{1\columnwidth}{!}{\includegraphics{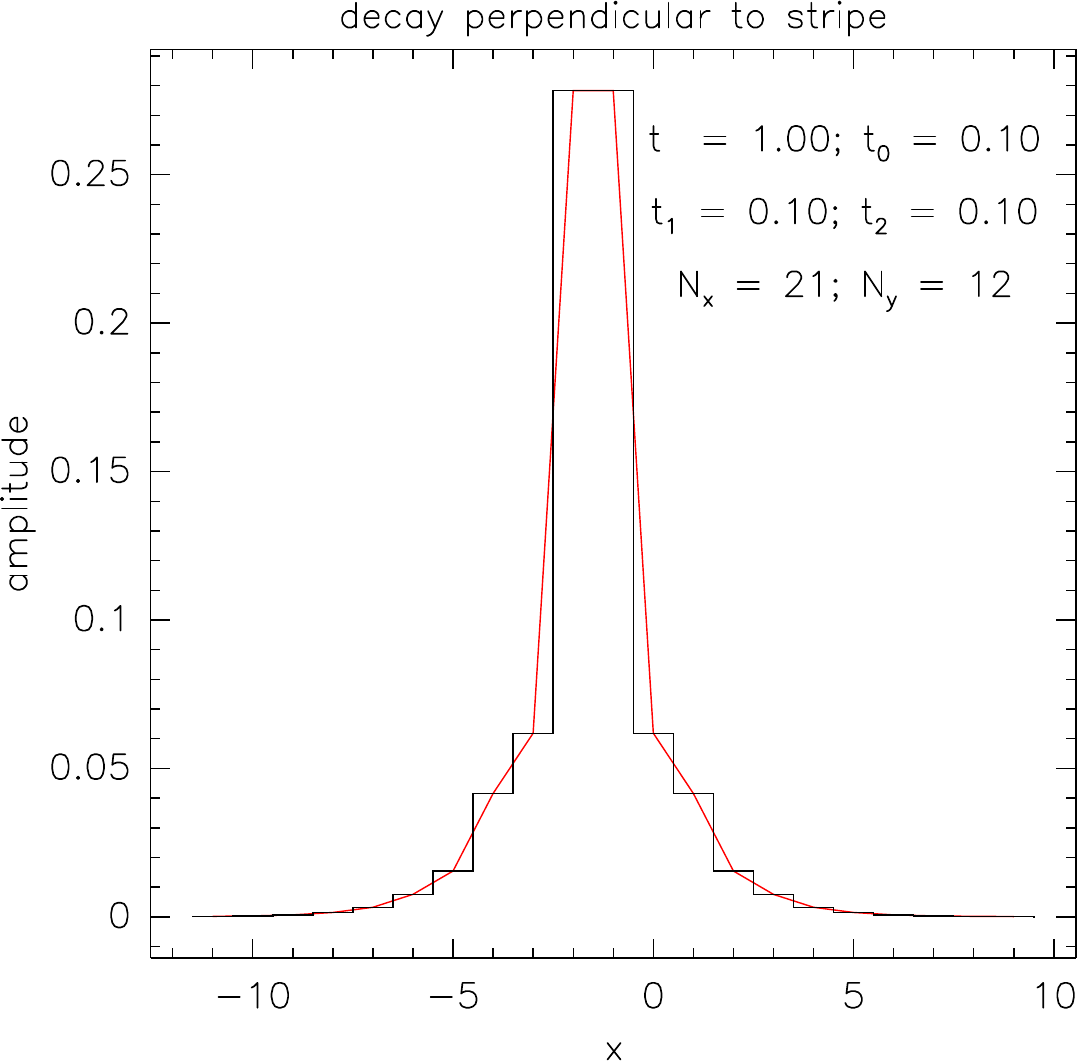}} \par}
\caption{Groundstate wavefunction perpendicular to a domain-wall for a single
hole in an antiferromagnet. This was calculated using the simple approximation
discussed in the maintext. The hole prefers to be on the stripe.
We find 
\label{figure: decay perp to stripe}
\protect\( |\psi |\approx |\frac{t_{0}}{t}|^{y}\protect \). The line
in the figure just connects the midpoints. $N_{x}$ and $N_{y}$
are the linear extents of the two dimensional
system and the hopping amplitudes $t_0$, $t_1$ and $t_2$ are
as defined in Fig. (3).}
\end{figure}

The reason is that within the antiferromagnet the holes can
only move in two steps longitudinally with an amplitude $t_0$ or
diagonally with an amplitude $t_2$ which are of the order of $J =
t^{2}/U \ll t$ (the uniniated reader is referred to 
Appendix(\ref{sub_prin}) for an exposure to the origin
of the exchange coupling $J$).
On the stripe the
hole can stay on the same sublattice by doing a single (instead of
double) step up or down with amplitude \( t \). As a consequence, the hole
has a much larger kinetic energy if it is situated on the stripe. 
Therefore we find that the hole automatically drifts toward the stripe: the
amplitude for the ground state wavefunction is maximal for sites on the stripe
and decays exponentially the further the sites are away from the strip
(see Fig. \ref{figure: decay perp to stripe}).
Holes 
are driven by an increase in kinetic energy (not exchange energy)
to the domain wall. We term this mechanism ``{\em dynamical confinement}''.
The word dynamical is used because it is the motion of the holes on
the stripe that lowers the energy.
The primary role of kinetic energy in favoring the stripe as a ground state
has been discussed many times before, e.g. \cite{pryadko,strings,erica}.
Nonetheless, the trivial (yet sizable) lowering of the kinetic energy allowed by hole motion along 
the ferromagnetic rungs on the domain wall seems to have gone unnoticed.

\begin{figure}
{\centering \resizebox*{1\columnwidth}{!}{\includegraphics{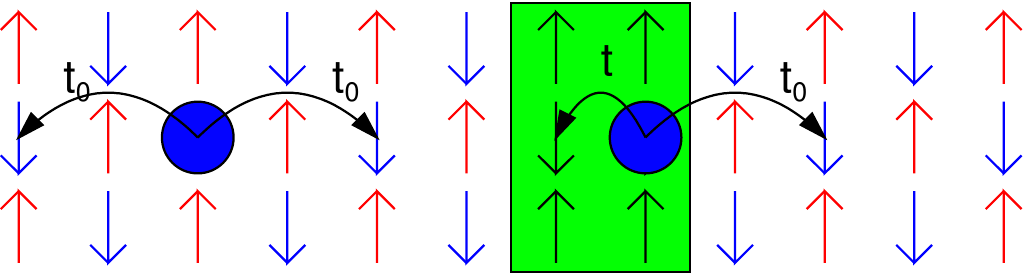}} \par}
\caption{One-dimensional example showing the reason for {\em dynamical confinement}.
Outside the stripe, the hole moves in steps of two with an amplitude
\protect\( t_{0}\protect \). Inside the stripe it can make a direct
hop with amplitude \protect\( t\protect \) with \protect\( t\gg t_{0}\protect \).
\label{figure: 1 dimensional hole confinement thing}}
\end{figure}

For the two dimensional problem of a single hole  
in the vicinity of a full domain wall embedded in a surrounding 
antiferromagnet (of the variant shown
in Fig. (\ref{figure: Hole in AF + domain wall})),
we may work in a very much reduced Hilbert
space. In effect, the strongly correlated
many particle system reduces
to a single particle problem defined on half of
the lattice. 
As there are, {\it ab initio}, four possible electronic states
on each site (empty, one electron with its spin up, 
an electron with spin down, and an up and down electronic
pair), the Hilbert space of an $N$ site 
system is, trivially, of dimension \( 4^{N} \).
Invoking the sublattice parity principle allows us to reduce the
Hilbert space to a mere size of \( N/2 \).
The Hubbard Hamiltonian is then reduced to a 
purely kinetic model, having an amplitude
$t$ for nearest neighbor direct hops within
the stripe, amplitudes $t_{0}$
and $t_{2}$ for longitudinal 
and diagonal two step hops (as depicted in Fig. 
\ref{figure: 1 dimensional hole confinement thing}
for the horizontal motions) within the antiferromagnet
and an amplitude $t_{1}$ for two step hopping within
the stripe along the axis of the ladder.

Before doing a more detailed analysis, let us first
give the reader an intuitive feeling 
of where we are heading.
Looking at Fig. (\ref{figure: Hole in AF + domain wall}), we note that 
a hole in the bulk disperses with an energy

\begin{eqnarray}
\epsilon_ {\mbox{bulk}}&=& \epsilon_{0}-2t_0 (\cos 2 k_{x} + \cos 2 k_{y}) \nonumber \\
	&& - 2t_{2} (\cos(k_{x}+ k_{y}) + \cos(k_{x}-k_{y})). 
\end{eqnarray}

Similarly, along the ladder, 

\begin{eqnarray}
\epsilon_{\mbox{stripe}} = \epsilon_{0} \pm t - 2 t_{1} \cos 2 k_{x},
\end{eqnarray}
where the second, $\pm t$, contribution denotes 
the energies of the bonding/antibonding states
along the rung.
Within this approximation, a sizable 
finite gap $\Delta\approx t$ separates the minima
of both dispersions. In our analysis,
it is this ${\cal{O}}(t)$ gap
which is the main driving
force for self consistent 
stripe formation having nothing direct
to do with string states and magnetic alleviation
energies (which are all of order ${\cal{O}}(J)$).

In reality, the hole can hop between the stripe
and its surrounding bulk: the stripe states
may be connected to the bulk. 
The low lying stripe states
are however much lower in energy than
their counterparts within 
the bulk. As a consequence, holes 
become localized on the stripe. 
As $t \gg t^{2}/U = {\cal{O}}(t_{0}) =
{\cal{O}}(t_{2})$, the on stripe dispersion is  
markedly lower in energy 
than its counterpart for motion
within the antiferromagnet. This leads
to an exponential decay of the lowest 
eigenstates states out of the stripe.

We have numerically diagonalized a 12 x 21 system
with a direct nearest neighbor hopping amplitude $t= 1$,
effective longitudinal and diagonal next nearest neighbor hopping
amplitudes $t_{0}= t_{2} = -0.1$ within the antiferromagnet,
and a next nearest neighbor amplitude along the axis of the stripe (ladder)
$t_{1} = -0.1$.  The groundstate wavefunction is, to good numerical
accuracy, given by 

\begin{eqnarray}
\psi (x,y) \approx |\frac{t_{0}}{t}|^{y}e^{ik_{x}x}.
\label{wave}
\end{eqnarray}

Fig. (\ref{figure:
decay perp to stripe}) depicts this wavefunction along the direction perpendicular to the
stripe. This figure is similar to Fig. 22 from Chernyshev {\em et. al.} 
\cite{strings} who calculated this from a $11 \times 7$ $t$-$J_z$ system employing the
numerical DMRG method. Our approach leads directly to this exponential confinement due to
the {\em dynamical confinement}, without the need for complicated numerical calculations.

We will now aim to give the reader an intuitive grip
on the physics 
along with the ability to easily
derive rigorous bounds on hole localization 
within the planar problem.

If we were to allow a hole on the 
particle to hop to a larger number
of neighbors within 
the bulk (i.e. to artificially
enhance kinetic motions and
lower confining
tendencies), then we
may map our system
into the dynamics 
of pseudo-spin 1/2 particles
in the presence
of a transverse magnetic
field. The mapping is as 
follows: let us mark 
each point within the 
bulk by its sublattice 
parity (up/down) number
as depicted in Fig.(\ref{figure: Hole in AF + domain wall}).
Let us now tile the plane into vertical
2 $\times$ 1 domino blocks lying with 
their long side parallel to the $y$ axis. Each domino
within the bulk contains one up and one down
site; the labeling of the up and 
down sites on the dominos lying along 
the rungs of the ladders may be done
arbitrarily. Next, let us envision replacing
each domino by a single spin-1/2 particle:
the number of the fictitious spin-1/2 particles
is equal to half of the number of sites
which we now label by $m$ and $n$.  For simplicity
we will now assume that all second order hops are of equal
magnitude $\sigma$.
Let us now consider the Hamiltonian 
\begin{equation}
\label{new}
H= - \sigma\sum _{\langle mn\rangle, d }(c_{m d }^{\dag }c_{n d}+c_{n
d }^{\dag }c_{m d }) - \sum _{n \in ladder}
\vec{h}_{ext} \cdot \vec{D}_{n},
\label{zim}
\end{equation}
with the fictitious external magnetic 
field $\vec{h}_{ext} = t \sigma_{x} \hat{e}_{x}$ oriented
along the transverse x-axis, and $\vec{D}_{n}$ the spin
of the psuedo-particle (domino) along 
the n-th rung of the stripe and $d$ is the
pseudo-spin polarization label. In the first term
$m$ and $n$ span the entire plane; the sum
is performed over all nearest neighbor
site $\langle mn\rangle$ and on 
four of the eight next nearest
neighbor sites (such
that all same sublattice
hoppings of the holes within
the bulk are accounted for).
For holes not far from 
the stripe or on it, the Hamiltonian of Eqn.(\ref{new}) 
introduces additional unphysical motions. These 
allow hoppings of the holes off and on the stripes which 
are disallowed-
such additional terms can only enhance delocalization
tendencies. The second term in Eqn.(\ref{new}) codes for the 
nearest neighbor up/down hops along the same rung- these 
alter a pseudo-spin up state to a down state
and vice versa.
We trivially observe that the low energy dynamics
corresponds to the motion of a $| \rightarrow \rangle$ 
particle polarized along
the applied transverse field 
sensing an attractive confining
potential of strength $t$ along the stripe. 
In physical terms, the psuedospin polarized 
state $| \rightarrow \rangle = 2^{-1/2}(|\uparrow \rangle +
|\downarrow \rangle)$ corresponds
to the symmetric low energy bonding
state (and its counterpart $| \leftarrow \rangle$
corresponds to the high energy antibonding 
state). The ground state is simply that of 
a polarized  $| \rightarrow \rangle$ particle 
(the symmetric bonding state) sensing a well of infinite extent
along one axis and having a finite extent ($2a$, with $a$ the lattice
unit) along the transverse direction. As the problem 
is translationally invariant along the ladder 
axis, the longitudinal momentum $k_{x}$ is a good quantum number.
The lowest lying wavefunction is translational invariant along
the infinite cylindrical axis ($k_{x}=0$) and, at long distances,
has the transverse profile  
of a particle of an effective mass $m_{eff} = 1/(2 \sigma)$ 
subject to the influence
of a potential well of depth $t$ and width $(2a)$.
This trivially leads to an exponentially
decaying amplitude in the direction transverse
to the stripe. As $t \gg J$, this confining
tendency is much more profound (and physical)
than most of the common magnetic bond ($J$) 
arguments prevalent in the literature. The reader
should bare in mind that our point
of departure- the sublattice
parity principle is correct
only as low energy scales 
(as compared to $J$). Nonetheless,
what drives localization
in our picture at
low energy scale are
the much more significant
kinetic effects which
outshadow the common
bad bond ($J$) 
counting arguments. 

The solution
to this potential well problem is standard.
Scaling back by a factor of two to the original (non-domino) coordinates along
the $y-$ axis, we find, asymptotically, 
\begin{eqnarray}
\psi(x,y) \approx  A \exp[- 2 \alpha |y|];~ |y| \gg a, 
\label{bound}
\end{eqnarray}
with the lowest lying bound state of Eqn.(\ref{zim}) 
is obtained by the solution 
of the transcendental 
equations 
\begin{eqnarray}
[\beta a \tan \beta a  = \alpha a]  ~\mbox{and}~ [a^{2}( \alpha^{2} + \beta^{2}) = \frac{1}{ \sigma}t] 
\nonumber \\~ ~\mbox{or}~~
[\beta a  \cot \beta a = - \alpha a] ~\mbox{and}~ [a^{2}(\alpha^{2}+\beta^{2}) =  \frac{1}{ \sigma}t]
\end{eqnarray}
having the largest value of $\alpha$.

Eqn.(\ref{bound}) gives an upper bound on the leakage
of the ``hole'' wavefunction out of the stripe.
If we set $\sigma= $ max $\{t_{0},t_{1},t_{2}\}$, 
we obtain a rigorous upper bound on $|\psi|$, at large
distances $|y|$, 
consistent with our numerical findings
(Eqn.(\ref{wave})). As the Hamiltonian of
Eqn.(\ref{zim}) contains additional unphysical
hopping processes from the 
ladder to its environment,
a localized state found
for the Hamiltonian of
Eqn.(\ref{zim}) implies
even more localized states
for the more restricted physical
problem.

We may similarly address the problem of
an arbitrary stripe configuration. In
general, adding additional hoppings
transforms the problem of the hole motion
within and/or near stripes
to a kinetic planar problem of
a hole coupled by Zeeman couplings
to magnetic fields piercing the plane 
along the stripe trajectories. A multitude
of viable self consistent minima of narrow
stripes of various 
geometries are found. Stripe dynamics is akin
to the motion of the fields (solenoids)  
piercing the plane; these, in turn
affect hole dynamics
by a Zeeman like effect. The evolution
of a hole-stripe system may be addressed
via such a self consistent scheme.

A related way of immediately deriving lower bounds on
the diffusion of the hole out of the stripe amounts to
looking at the transverse cross-section
of the stripe as depicted in 
Fig. (\ref{figure: 1 dimensional hole
confinement thing}), and examining
single hole motions. Solving the Schrodinger
equation for this one dimensional system, we immediately
obtain a localized bonding state of the 
lowest energy. As here fewer hops are
accounted for than in the original
physical problem, the solution 
serves as a lower bound on ``hole''
wavefunction leakage out of the
stripe. When fused with the 
previous upper bounds 
from the pseudo-spin system,
we immediately obtain both
lower and upper exponential
bounds on asymptotic hole 
propagation out of the stripe.

In this scheme, splintering the sites according to their
sublattice numbers (equivalent to
the magnetic field sensed by a 
spin of definite polarization if it 
is placed at various locations
throughout the lattice) introduces hopping 
parameters for the low energy dynamics
with no potential energy in sight. 
As shown by our analysis, 
magnetic alleviation effects
are not imperative for achieving this
dynamical confinement (even when direct magnetic
effects are removed we numerically attain exponential
localization with little noticeable
change). We reiterate that
we are examining the reduced Hilbert
space where no magnetic string 
states exist from the 
outset.

In our approach, we first create an antiferromagnetic spin structure
with domain walls and then dope the system. Then,
if the spin structure is strong enough, the holes will migrate to
the walls and will form a charge structure. Hopefully we have
convinced the reader that sublattice parity order can 
indeed drive stripe formation. The energy scale associated
with this discrete sublattice parity ($Z_{2}$) order \cite{zohar} 
can indeed be quite high: the persistence 
of incommensurate peaks up to at least the high energy resonance peak 
in YBCO might be interpreted as an indication that the 
stripe persists as a domain wall
up to very high energy scales \cite{resonance}.
Experimentally, charge order is a far greater
robust driving force for stripe formation than 
spin order. The role of sublattice 
parity (albeit its high
energy scales) is not clear 
at the time of writing. The reader should consider 
our argument as a self-consistent one in which the creation
of the no-hole domain-wall first is a theoretical device that simplifies
theoretical discussion. 
Summarizing, we have seen that once there is a ferromagnetic seam in
a very strong antiferromagnet, single holes will automatically move
to this seam and be exponentially localized onto them. In the remaining 
part of this article we will see what the consequences are of
this result for more than one hole on the stripe.
The localization of the holes is in accord with 
NQR measurements, e.g., \cite{savinkov}.

\section{Staggered Ladder systems}
\label{LaDDeR}

We have seen that by dynamical confinement we may consider
an antiferromagnetic domain wall as a two-leg ladder with staggered boundary
conditions. So, once again
we can restrict our Hilbert space.
Ladder systems have been extensively studied throughout the
years. 
Hundreds of
works on standard (unstaggered) ladder systems
have been carried. 
For a well-known review see Dagotto and Rice \cite{DR}.

In our case, the influence of the 
antiferromagnet surrounding the ladder
is still there. The surrounding antiferromagnet 
effectively gives rise to staggered boundary conditions.
In the up and coming, we will look at two-leg ladders with staggered 
fields mimicking the surrounding antiferromagnet. The
staggered fields endow the ladder with 
a sublattice parity structure. 
Unfortunately, to date, staggered ladders have
not been investigated intensively. 

Krotov, Lee, and Balatsky \cite{KLBafboundary}
examined staggered ladders for the Hubbard model at small \( U \).
In this limit, one starts with a non-interacting system and then derives
the renormalization group equations. 
Staggered spin ladders without any holes have been studied 
by Wang {\em et. al} \cite{Wang}.
However, staggered ladder systems with holes in the large $U$ limit have not
yet been addressed before as far as we know. 
We will now analyze staggered two leg 
ladders in a rather pedestrian manner. 
We verified the perturbative results that we will cite 
by explicit numerical computations.

\section{The Single electron on an empty staggered ladder}
\label{empty-one}

To understand the effect of the boundary conditions on the movement
of electrons and holes on the ladder we first examine a single electron
on an empty ladder immersed in a staggered 
magnetic field. This is, of course, a hypothetical situation, but
it is nevertheless a good starting point for our discussion. If the
electron is on the correct sublattice (i.e. if
its spin polarization is opposite
to that of the applied 
staggered magnetic field) then its magnetic energy will be
\( \epsilon_{0}=-J \).
Being on the wrong sublattice leads to a magnetic 
energy penalty of \( \epsilon_{1}= +J \).

\begin{figure}
{\centering \resizebox*{1\columnwidth}{!}{\includegraphics{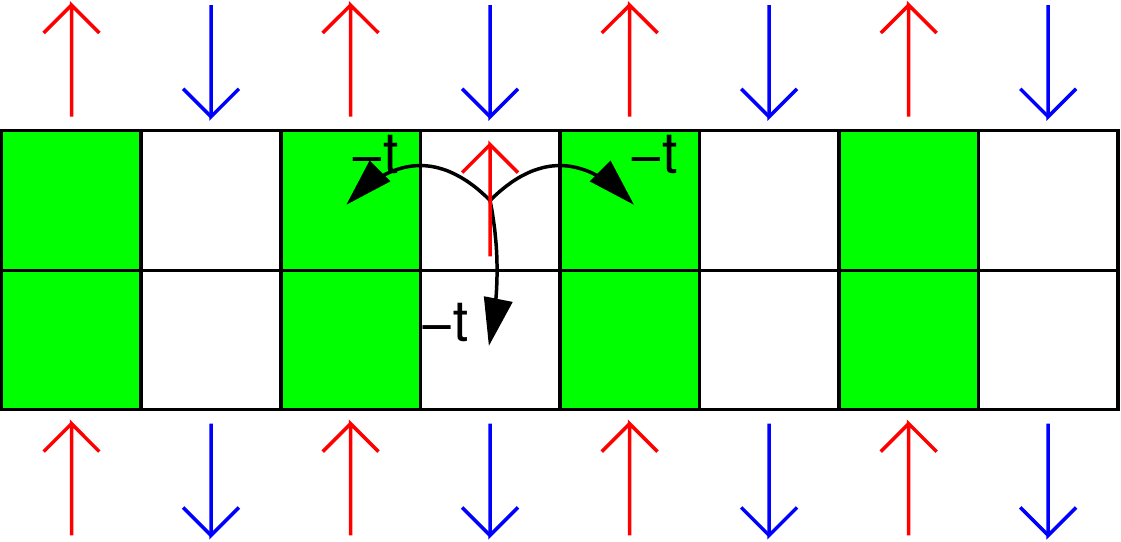}} \par}
\caption{Single electron on an empty ladder. In principle the electron can
hop to all its nearest neighbors with an amplitude \protect\( t\protect \).
However, the ladder is embedded in an antiferromagnetic background
that influences the electron on the ladder.}
\label{single_electron}
\end{figure}

This gives rise to staggered potentials on the ladder. 
The resulting Hamiltonian for the system as shown in Fig. (\ref{single_electron}),
reads

\begin{equation}
\label{equation: Hamiltonian matrix}
H=\left( \begin{array}{cccccccccc}
\epsilon _{1} & -t & -t &  &  &  &  &  &  & \\
-t & \epsilon _{1} & 0 & -t &  &  &  &  &  & \\
-t & 0 & \epsilon _{0} & -t & -t &  &  &  & 0 & \\
 & -t & -t & \epsilon _{0} & 0 & -t &  &  &  & \\
 &  & -t & 0 & \epsilon _{1} & -t & -t &  &  & \\
 &  &  & -t & -t & \epsilon _{1} & 0 & -t &  & \\
 &  &  &  & -t & 0 & \epsilon _{0} & -t & -t & \\
 & 0 &  &  &  & -t & -t & \epsilon _{0} & 0 & -t\\
 &  &  &  &  &  & -t & 0 & \epsilon _{1} & -t\\
 &  &  &  &  &  &  & -t & -t & \epsilon _{1}
\end{array}\right).
\end{equation}

Here we express the Hamiltonian in terms of ordered real-space basis states
where we place the upper and lower sites of each rung adjacent to each
other. An electron at any site can always hop to three other sites. 
Because this is a single particle problem with a translational invariant 
potential (with unit cell of size two), the
eigenfunctions of this Hamiltonian can exactly be determined. They are 
exactly given by 

\begin{equation}
 \psi =(B,\pm B,Ae^{ik},\pm Ae^{ik},Be^{2ik},\pm Be^{2ik},Ae^{3ik},...).
\end{equation}

Here \( A \) and \( B \) are determined by

\begin{eqnarray}
E_{k} & = & \epsilon _{0}\pm t-2t\frac{B}{A}\cos k \nonumber \\
E_{k} & = & \epsilon _{1}\pm t-2t\frac{A}{B}\cos k.
\end{eqnarray}

The energy can immediately be obtained from 

\begin{equation}
\left\Vert \begin{array}{cc}
\epsilon _{0}\pm t-E_k & -2t\cos k\\
-2t\cos k & \epsilon _{1}\pm t-E_k
\end{array}\right\Vert =0,
\end{equation}

leading to

\begin{eqnarray}
\label{equEK}
E_{k}&=&\pm t\pm \sqrt{J^{2}+(2t\cos k)^{2}} \nonumber \\
&=&\pm t\pm\sqrt{J^2+2t^2+2t^2\cos 2k}.
\end{eqnarray}

This dispersion illustrates that, due
to the staggered boundary conditions, the
electron effectively moves in steps of two. 
Fourier transforming the energy shows that
because of the symmetry between $k$ and $\pi-k$, 
only terms which have an even number of steps from the
starting point may have an amplitude unequal to zero: odd,
sublattice parity interchanging, hops are banned. 
In more physical terms, this trivially
corresponds to the halving of the period (in $k$)
in the dispersion $E_{k}$.

\begin{figure}
{\centering \resizebox*{!}{4.45cm}{\includegraphics{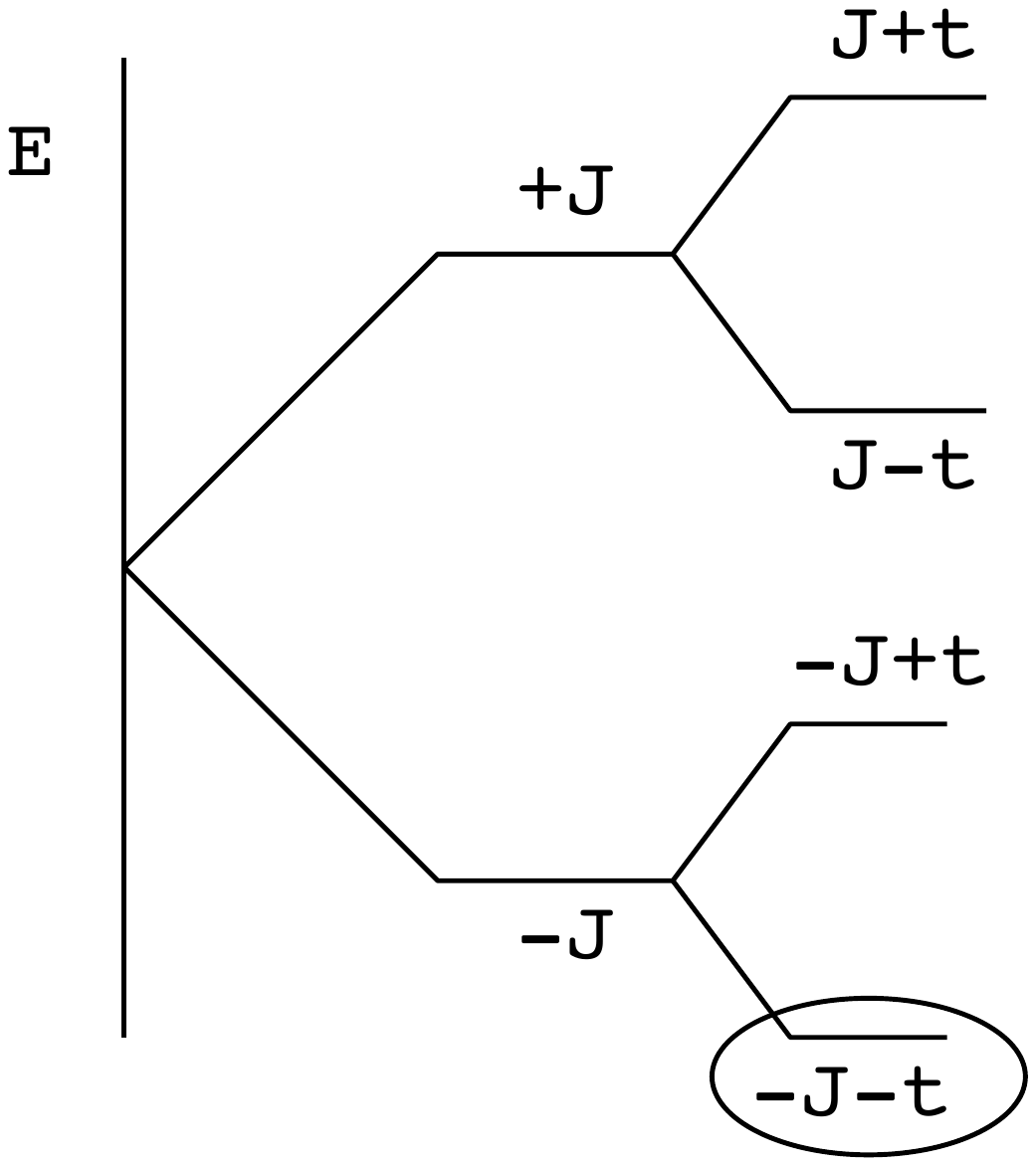}}}
\hspace{1cm}
{\centering \resizebox*{!}{4cm}{\includegraphics{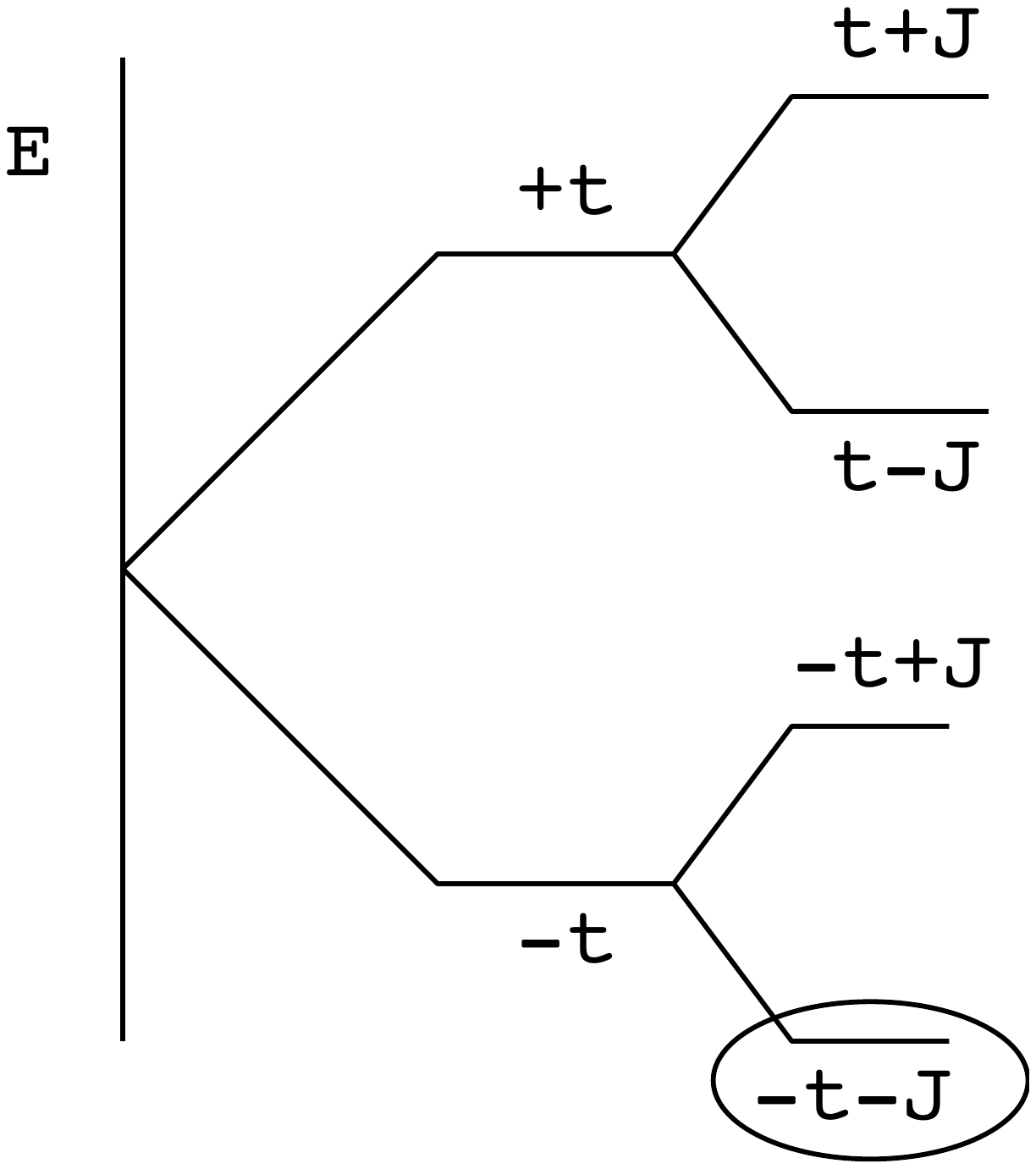}} \par}
\caption{Energy diagram for a single electron on an empty staggered 
ladder. If the staggered
potential is large (large \protect\( J\protect \), see left panel), then the splitting
set by $J$ will be more significant than that associated with \protect\( t.\protect \) Within the lowest energy sector,
there are only \protect\( N/4\protect \) states. These are the same
states as for a single one dimensional model where an electron moves in steps of two.
The corresponding schematic for $J\ll t$ (having the same lowest energy) is depicted on the right. 
\label{figure: energylevel diagram for one spin on ladder}
}
\end{figure}

Eqn. (\ref{equEK})
shows that, for unphysically large \( J \gg t\) (a strong 
influence of the surrounding two-dimensional
antiferromagnet), the Hilbert space splits up in four different sectors.
First, there is a splitting because of the boundary conditions:
half of the sites have the electron on the right sublattice, which
leads to a low energy of \( -J \). The other half have the
hole on the wrong sublattice, with an energy of \( +J \).
Because the upper and lower leg of the ladder
are exactly equivalent, the electron will slosh back and forth
between the upper and lower legs. This, in turn, leads to bonding/antibonding
linear combinations of the upper and lower sites along each
rung, further splintering the Hilbert space into
two additional subsectors. The lowest sector, which contains 
$N/4$ basis states, consists of wavefunctions with
the electron on the right sublattice 
(e.g. a spin up polarized electron on the 
up sublattice) with the symmetric linear combination (bonding)
of upper and lower sites. 

In the large $J$ limit, Eqn. (\ref{equEK}) simplifies to

\begin{equation}
E_{k}=-J-t-2\frac{t^{2}}{2J}-2\frac{t^{2}}{2J}\cos 2k.
\end{equation}

Not surprisingly, we see that this energy
corresponds to the electron populating the correct 
sublattice (\( -J) \) and being smeared
along the two rungs in a symmetric bonding fashion (\( -t \)). 
We can easily read off that the effective
hopping is in steps of two from the \( \cos 2k \) term. Because we
are in the large \( J \) limit, the hopping amplitude in second order
perturbation theory is given by \( \frac{t^{2}}{2J} \). We have to
hop twice (\( t^{2} \)) over an intermediate state with energy \( 2J \).
The extra contribution \( -2\frac{t^{2}}{2J} \)to the energy comes
from virtual excitations where the spin moves one position to the
left or right and immediately returns back. There are two possible
ways of doing this, and the amplitude again is \( \frac{t^{2}}{2J} \).
Thus, we can intuitively understand every term in this limit.

There is a gap of \( 2t \) separating the lowest Hubbard band
from the antibonding states. Because \( 2t \) is a very
large energy scale (approximately 0.5 eV, or 5000 K), we can neglect the
influence of the other sectors and may restrict our attention
only to the lowest sector.

\begin{figure}
{\centering \resizebox*{0.7\columnwidth}{!}{\includegraphics{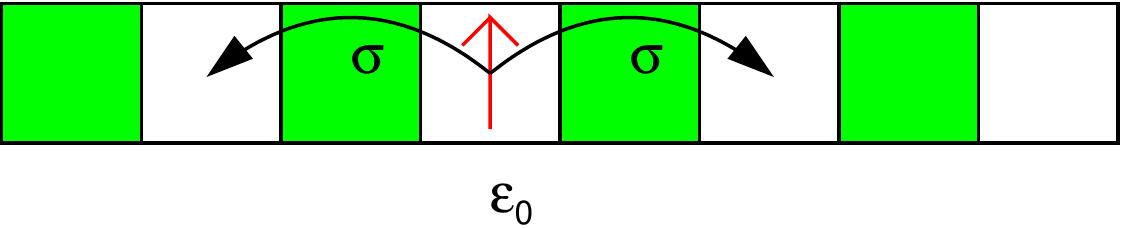}} \par}
\caption{Effectively, the electron moves on a one-dimensional chain in steps of
two. There are only two effective parameters: the on-site energy 
\protect\( \epsilon _{0}\protect \)
and the two site hopping amplitude 
\protect\( \sigma \protect \).\label{figure: spin on 1 dim}}
\end{figure}

Our discussion above hinged on the assumption that 
\( J \) is very big (\( J\gg t \)). This assumption is not satisfied
in the physically relevant region wherein
\( J\approx 0.25t \). In the physically relevant regime, we still have a splitting of the
Hilbert space into four sectors (see 
Fig. (\ref{figure: energylevel diagram for one spin on ladder})).
The lowest Hilbert sector is still the same one as the one for the
large \( J \) limit. 
For small values of $J$ compared to $t$, the groundstate properties are still dominated by the $-J$, $-t$ sector.
However, the low lying  excitations out of the lowest
sector are now to the sector where the electron goes
to the wrong sublattice (a single hop) instead of going from a bonding
to an antibonding state. Such hops will lead to string states. Nevertheless, as long as the 
width of the lowest
Hubbard band is small compared to \( 2J \) (which itself is also a large
energy, of the order of 2400K), we can still think in terms of bonding
states on the correct sublattice. Within the lowest Hubbard sector,
the electron will only be on the even lattice sites. Effectively,
only second order hops are present. 
Because of the hybridization
of up and down sites on a rung, it is sufficient to consider 
the one dimensional
model shown in Fig. (\ref{figure: spin on 1 dim}).

Just as for the single hole
in the two dimensional antiferromagnet, 
discussed in Appendix (\ref{sub_prin}), 
a lone electron
on the empty ladder moves as an effective quasiparticle, with a sublattice hopping
amplitude $\sigma$.

\section{One hole on a full Staggered ladder}
\label{full-one}
We now examine one hole in an otherwise full ladder
immersed in a staggered external magnetic field. The single spin problem was 
very simple to solve analytically given our assumptions. The 
single hole problem is exceedingly
more difficult because it is a strongly interacting many
body problem. For a system of only $6\times 2$ sites, there are already
approximately $800,000$ states! Systems of this size and larger
may be addressed by employing the Lanczos method of diagonalization, which
only finds the lowest eigenstates of a matrix.
We employed this method for systems of up to size $8\times 2$. The 
complicated results
that follow from this numerical study can be understood quite easily for
large values of $J$. Though not the physically relevant regime, 
just as in Section (VII), both for $J>t$ and $J<t$, the ground state
is in the same sector of Hilbert space.

For large values of $J$, the basic properties are once again those of a 
single quasiparticle. From the outside, we again assume that the hole cannot leave
the stripe. An electron next to the hole can move to the position
of the hole, leading to a string-state with one wrong spin.
The hole can also move up and down. And it can move in steps of two
to the left and right. In addition, we can have spin flips on the ladder.
However, just as in the two-dimensional antiferromagnet we neglect
the spin-flip processes. For large values of $J$ we find once again a number 
of Hubbard sectors. The lowest sector has N/2 states where 
the hole is on the right sublattice and an energy
of approximately \( -2(N-1)J \). Above this lowest Hubbard sector
we have the Hubbard sector with a string state of length one. There
are 2N states in this sector. The gap between this sector and the
lowest sector above it is \( \Delta =4J\approx 4800K \). 
We could ingnore this sector if this gap is large
with respect to the internal splitting of the lowest Hubbard sector.
This bandwidth is given by \( 4t'' \) with \( t''=\frac{t^{2}}{U+4J}\ll t \).
For \( U\approx 8t \) we have \( t''\approx 0.1t \):
both $J$ and $t$ have approximately the same amplitude.
However, for the groundstate properties, the relevant properties are
those of the lowest a single hole moving in steps of two.
 The band, shifted down by \( t^{2}/2J \) by  
virtual hopping, has the dispersion of an inverted cosine. The gap 
\( \Delta =t \), separating the on-stripe states from those in the 
surrounding antiferromagnet, is large with respect to \( t'' \).
The spins adjust to 
surrounding antiferromagnetic. In the $U,J \gg 1$ limit,

\begin{figure}
{\centering \resizebox*{0.7\columnwidth}{!}{\includegraphics{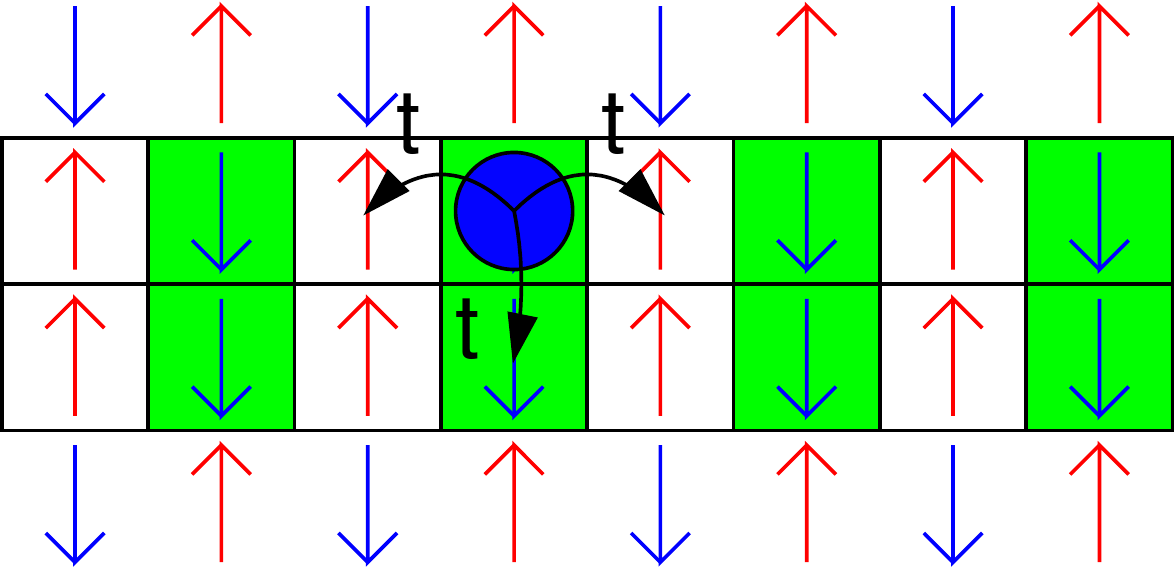}} \par}
\caption{We assume a single hole can move on the stripe and cannot leave the
stripe. The surrounding antiferromagnet leads to staggered boundary
conditions.}
\end{figure}

\begin{eqnarray}
\epsilon _{0} & = & -J-t-2\frac{t^{2}}{2J}, \nonumber \\ 
\sigma  & = & 2\frac{t^{2}}{U+2J}.
\end{eqnarray}

The motion of a single hole on a stripe and its motion in 
a two dimensional antiferromagnet is identical. 
The only difference between the parameter sets, is sparked by the presence of the direct hopping term 
$t$ allowed within the topological domain wall.
This is in contrast to one dimensional effective theories that assume 
that the hole effectively becomes a holon on a stripe (see 
for instance Tchernyshyov and Pryadko \cite{pryadko}). Our hole is not a 
holon in the sense that it will remember if it was injected for a spin up 
or a spin down electron: they move on diffent sublattices.

We see that we can consider a single hole on a stripe as a quasiparticle,
just as a single electron on an empty lattice. They can both be characterized
by (different) values for \( \epsilon _{0} \) and \( \sigma  \).
The main difference between a hole on a full ladder and an electron on
an empty ladder is that the hole moves more slowly, because
it can only move through double occupied intermediate
states, which costs a lot of energy. Therefore, \( \sigma  \) is small for a hole, while it is
large for an electron on an empty lattice. 
Our main conclusion is that a single hole and a single electron are
effectively identical.

\begin{figure}
{\centering \resizebox*{0.7\columnwidth}{!}{\includegraphics{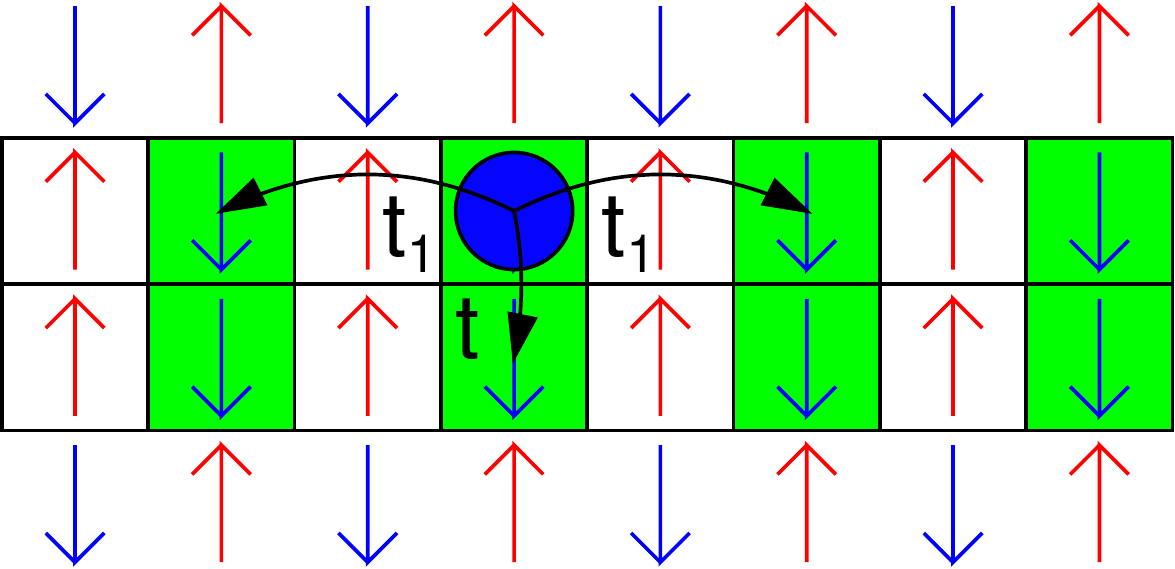}} \par}
\caption{For a one dimensional ladder with strong staggered potentials on
the boundaries, a single hole effectively moves in steps of two. }
\end{figure}

\section{Two electrons on an empty staggered ladder}
\label{2-empty}

\begin{figure}
{\centering \resizebox*{0.7\columnwidth}{!}{\includegraphics{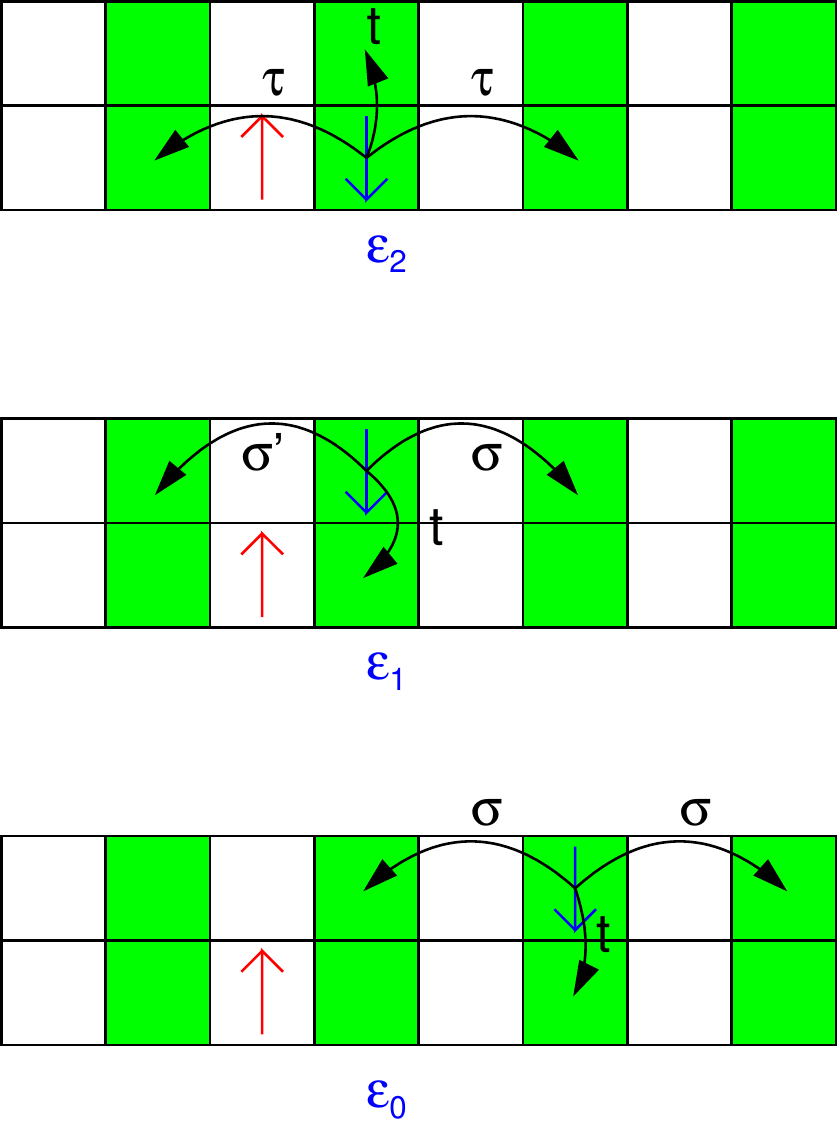}} \par}
\caption{Different possible configurations of two spins on a
staggered two leg ladder.}
\label{figure: geo on 1 dim}
\end{figure}

The equivalence between holes and electrons is no longer true for
more than one particle. We will show that the low energy properties
for two electrons are different from those of two holes.
We first consider the problem of two electrons with opposite spin on
an empty ladder. Various possible 
geometries are sketched in Fig. (\ref{figure: geo on 1 dim}). 
Because only the relative distance is important, we keep the position
of the up spin fixed.

The two electrons can be on the same or opposite legs. Six
amplitudes: \( \epsilon _{0},\epsilon _{1},\epsilon _{2},\sigma ,\sigma ' \)
and \( \tau  \),
defined in Fig. (\ref{figure: geo on 1 dim}),
are of relevance. 
We notice that the hopping amplitude for the
electrons when they are far apart ($\sigma$) is much larger than the hopping amplitude
when they are next to each other ($\tau$). If they are far apart,
they do not notice each other. Two electrons 
have less kinetic freedom if they are next to each other. In
order to move passed each other, they have to go through a doubly occupied
state.

In the \( U\gg J \) limit, lowest order perturbation theory yields

\begin{eqnarray}
\epsilon _{1}=\epsilon _{0} & = & -2J-4\frac{t^{2}}{2J}, \nonumber \\
\epsilon _{2} & = & -2J-2(\frac{t^{2}}{2J}+\frac{t^{2}}{2J+U}), \nonumber \\
\sigma  & = & \frac{t^{2}}{2J}, \\
\tau  & = & \frac{t^{2}}{U+2J}.  \nonumber \\
\end{eqnarray}

\begin{figure}
{\centering \resizebox*{0.8\columnwidth}{!}{\includegraphics{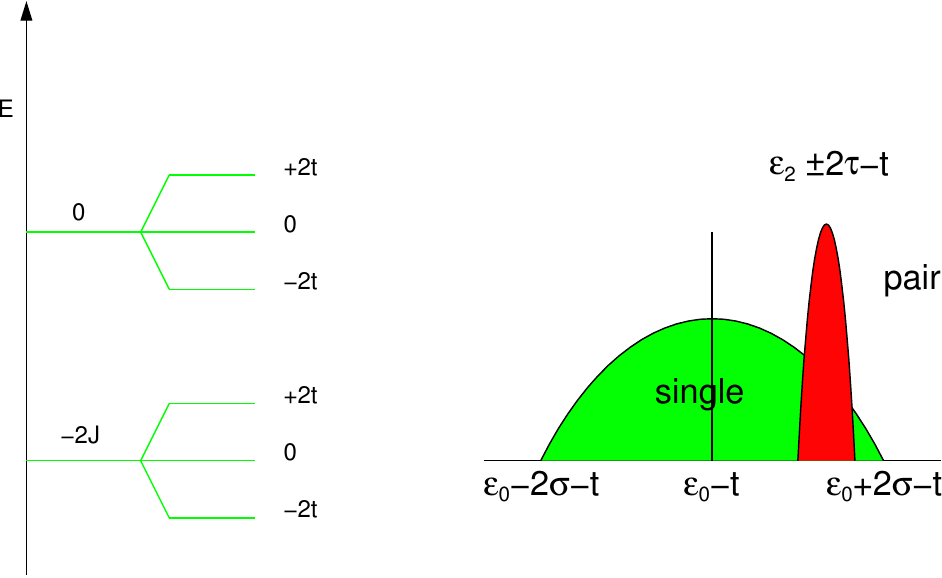}} \par}
\caption{Hubbard sectors for two spins on an empty ladder. The lowest Hubbard
sector contains both single electrons and pairs of electrons. However,
the groundstate consists of single electrons. Pairs are only formed
at a higher energy in the lowest Hubbard sector.}
\label{figure: two electron on ladder}
\end{figure}

The lowest Hubbard sector contains
both single electrons and pairs of electrons. However, we have seen
that single electrons have a larger kinetic energy than pairs. 

Therefore, the ground state consists of electrons
that are as far apart as possible. 
Fig. (\ref{figure: two electron on ladder}) shows a schematic picture for the
density of states. Only at the higher energies of
the lowest Hubbard band do we find pairs. The pairs have
a small bandwidth because of the small hopping amplitude.
Once again we find various sectors as shown in 
Fig. (\ref{figure: two electron on ladder}).

\section{Two holes in a full staggered ladder}
\label{2-full}

To see the difference between electrons and holes, we now look at two holes
in a filled ladder.
When far apart, two holes (quasiparticles) do not notice each other. 
Just as for two electrons on an empty
lattice, there are two states in which they strongly influence
each other. The main question we wish to address is whether
two holes come together and form a real-space pair, or 
if they tend to be as far apart as possible.

\begin{figure}
{\centering \resizebox*{1\columnwidth}{!}{\includegraphics{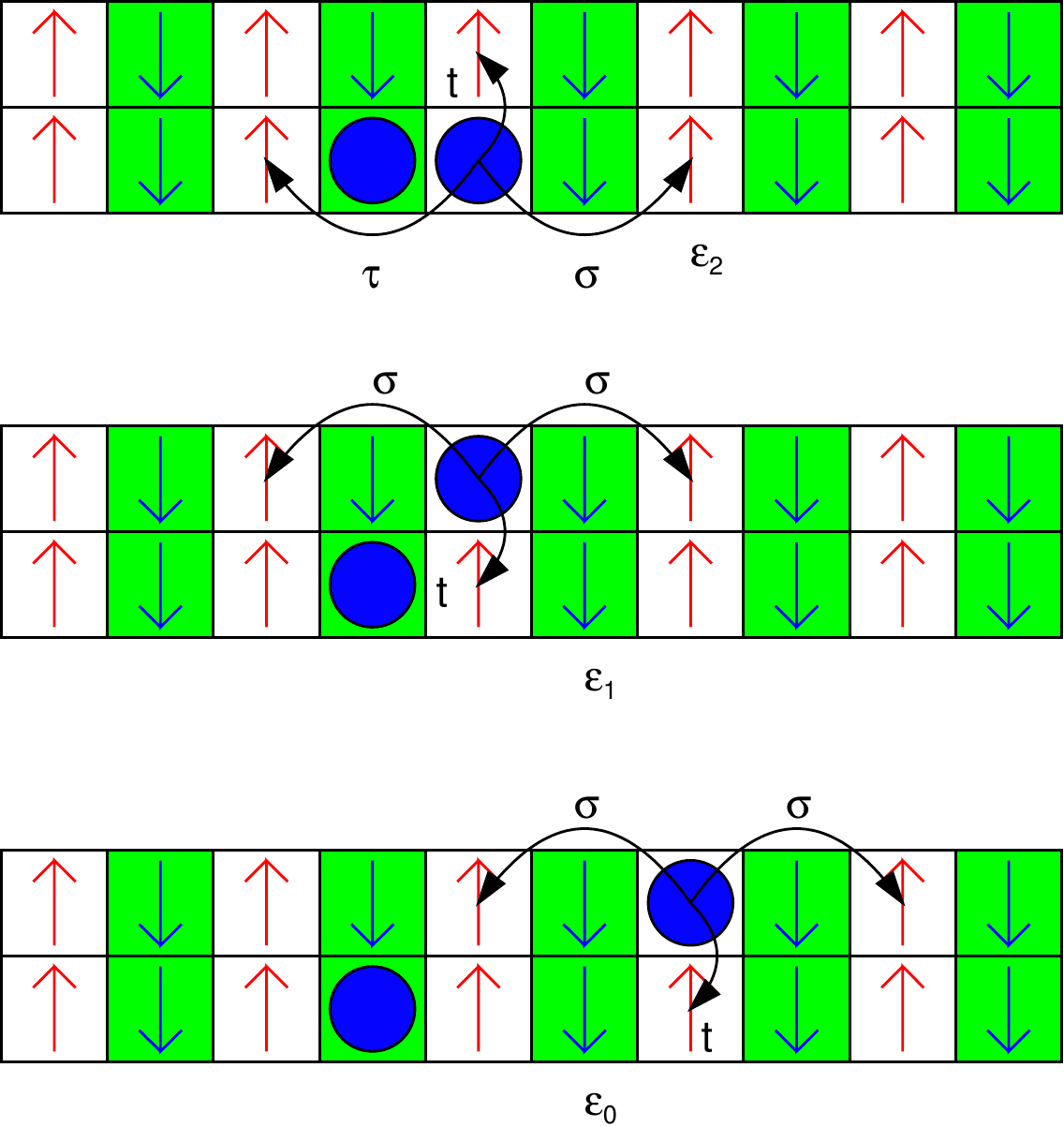}} \par}
\caption{
Different configurations of two holes on a stripe with the relevant on site
energies and hopping amplitudes. 
\label{figure: two holes moving indep on two-leg ladder}}
\end{figure}

If the two holes are far apart,
they will move independently with an amplitude \( \sigma =\frac{t^{2}}{U+2J} \)
(see Fig. (\ref{figure: two holes moving indep on two-leg ladder})).
However, this is different if they are sitting next to each other. In this case no double 
occupied state is needed. Therefore the hopping amplitude \( \tau =\frac{t^{2}}{2J} \), thus \( \tau \gg \sigma  \).

\begin{figure}
{\centering \resizebox*{1\columnwidth}{!}{\includegraphics{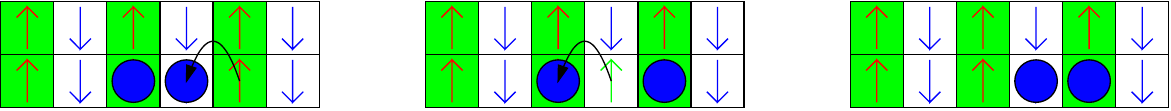}} \par}
\caption{Two holes can move together without creating an intermediate double occupied state. 
Therefore, holes want to form real-space
pairs. \label{hopping of a hole-pair}
A hole pair can move if a single
electron makes two hops. }
\end{figure}

If the two holes are on the same leg, sitting next to each other (see
Fig. (\ref{hopping of a hole-pair})),
than the two holes can move together with a much larger
amplitude than if they move alone. 

In the left subfigure, a spin up electron moves
left. The intermediate state in the middle figure has a higher energy
of the order of $J$. From this position, the electron with spin up can move back to
the starting position, or it can move on to end up as shown in the right
figure.
Once it is there, the hole-pair has effectively moved one site to the right. This
figure only illustrates one possible hopping sequence, there are also
other possible ways of hopping. 
For discussions of
pairing one should be careful with fermionic
minus signs sparked by the interchange
of identical spins once a pair
is made to go around \cite{trugman,erica}.
For the movement of a single hole,
there is an intermediate state with an energy of \( U \). However, if the
two holes are next to each other then the intermediate state will not
have a doubly occupied site, so therefore this energy is only of the
order of \( 4J \). Therefore, \( \sigma ' \) is much large than
\( \sigma  \). This leads to the formation of pairs on the stripe.

For large \( U \) and \( J \), employing the same 
convention as before,
we find in perturbation theory,

\begin{eqnarray}
\epsilon _{0} & = & -(N-2)J-4\frac{t^{2}}{2J}-(2N-8)\frac{t^{2}}{U+2J} \nonumber \\
\epsilon _{1} & = & -(N-2)J-4\frac{t^{2}}{2J}-(2N-8)\frac{t^{2}}{U+2J}  \\
\epsilon _{2} & = & -(N-2)J-2\frac{t^{2}}{2J}-(2N-6)\frac{t^{2}}{U+2J} \nonumber,
\label{energies}
\end{eqnarray}
with $N = 2L$ the total number of sites
on the two leg ladder. 

\begin{figure}
{\centering \resizebox*{1\columnwidth}{!}{\includegraphics{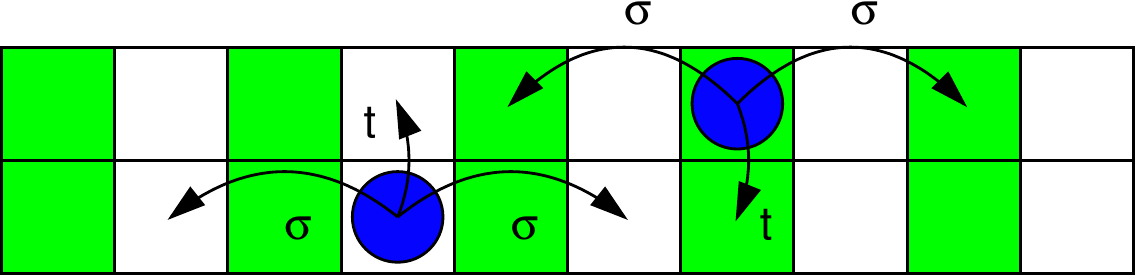}} \par}
\caption{We can neglect the spins on the ladder because they cannot change.
This is caused by the boundary conditions.}
\end{figure}

\begin{figure}
{\centering \resizebox*{1\columnwidth}{!}{\includegraphics{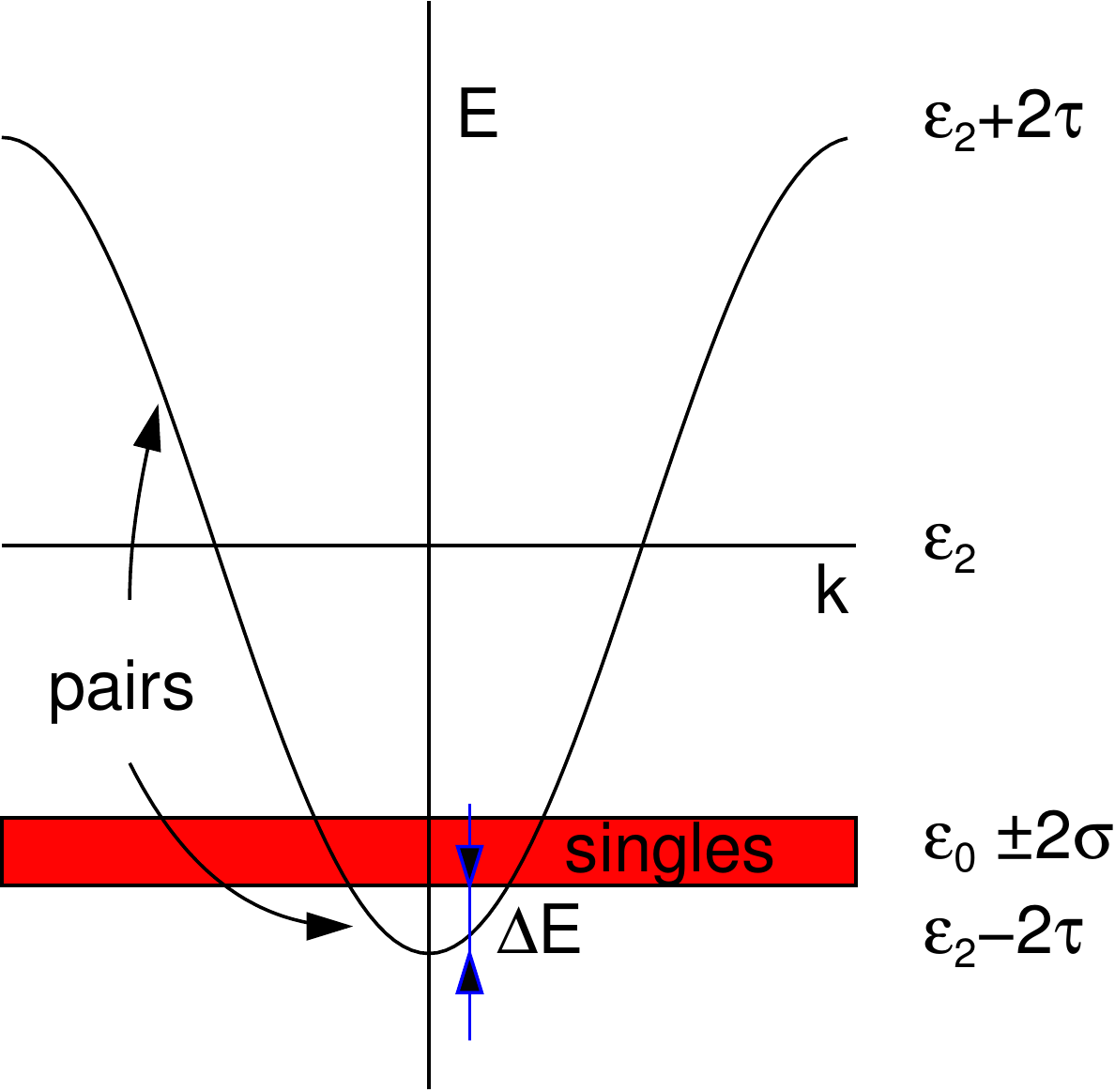}} \par}
\caption{Theoretical dispersion relation for two holes in a bond-centered
stripe.The groundstate consists of pairs.}
\label{dispersion}
\end{figure}

The motion of the pairs on a ladder and the motion of single holes is coupled: both are present
in the lowest Hubbard sector. If they form disjoint Hilbert spaces, we will find  \( E=\epsilon_{2}-2\tau \cos k \) for the real space pair. 
For two single holes, \( E=\epsilon_{0}-2\sigma \cos k \), within perturbation theory.
Because $\sigma$ is very small, the spectrum for single 
hole states does not exhibit much dispersion.
Taking the energy values from Eqn. (\ref{energies}), the dispersion for two holes on a stripe
with staggered potentials is suspected to look like that shown in Fig. (\ref{dispersion}).

The big difference between holes and electrons is that for two electrons the separated electrons
have a large bandwidth and the pairs of electrons are a high energy state with a small dispersion.
For holes this is reversed: the pairs have the largest bandwidth, while the separated electrons
form a low energy excitation (string states) with a small dispersion.

The lowest state for the pair cosine band is located at $\epsilon_2-2\tau$ and the lowest state 
for the separated holes is located at $\epsilon_0-2\sigma$. In
perturbation
theory,
both energies amount to
\begin{eqnarray}
-(N-2)J-4\frac{y^2}{2J}-(2N-8)\frac{t^2}{U+2J}.
\end{eqnarray}

However, here we assumed the pairs and separated holes to be independent. But where they overlap in energy, 
which in this case is the bottom of the band, the wavefunctions will hybridize.

If we think of a hybridization of up and down sites
along the rungs, then no possible interchange can be 
performed between the holes leading to fermionic
minus signs that elevate, rather
than depress, the pairing energy \cite{erica}.

\section{Lanczos calculations}

Lowest order perturbation theory for two holes on a staggered ladder leads to
the dispersion shown in Fig. (\ref{dispersion}). The ground state properties
cannot be easily determined because of the overlap between separate hole and
pair states. To test this situation numerically, we have performed standard
Lanczos calculations for the Hubbard model endowed with staggered
boundary conditions on a twelve site single chain
with staggered boundary conditions. Solving this problem for large
two leg ladders is inhibited by the large Hilbert space. The maximum size 
which we are able to examine numerically is a \( 6\times 2 \)
system for which there are \( \approx 650,000 \) states. Such a system
is too small to observe single hole states and
pairs accurately. Because of the strong bonding combination between the upper and
lower leg of the ladder, the reduction to a one dimensional line is physically justified.
The results of the Lanczos calculation show that the ground state has the largest
amplitude for pair states with only a small admixture of single hole states.
Immediately above the ground state, the excited states are predominantly single hole states.
There is a finite energy gap separating the ground state and these excited states.
Because of the small lattice size, we cannot determine the value of
the energy gap $\Delta E$.

Another approach is to assume that  \( \epsilon _{0},\epsilon _{2} \), \( \sigma  \),
and \( \tau  \) are adjustable parameters. In that case, we can perform a simplified
calculation on a $20 \times 2$ ladder having
twenty spin up, and twenty spin down sites. In the aftermath, this leads to a $400\times400$ matrix
that can be easily diagonalized. Using this model it is relatively easy to choose
the effective parameters such that the cosine band for the pairs has a lower energy 
than the states where we have single holes far apart. 
Our numerical results for the Lanczos method suggest that the gap \( \Delta E \) 
separating pair states and single particle states is sufficiently large for the groundstate 
to consist mostly of pair states.

Nonetheless, as the gap separating the single 
hole and pair states is 
small, the pairing physics
is, a priori, susceptible to the 
addition of longer range interactions 
and hopping interactions. Such effects
could easily tip the balance between
the single hole and pair
states (or enhance pairing tendencies).

\section{Putting all of the pieces together: extended lattice states}
\label{DMRRRG}

We found that if we assume that, somehow, the antiferromagnet
has a ferromagnetic domain wall, holes will automatically localize
on this topological line. Furthermore, we looked at the movement
of a single hole on this staggered ladder and found that it behaves
as a single quasiparticle that moves along the ladder in steps of two.
Two electrons on an empty lattice try to be as far apart as possible.
Two holes on an otherwise half filled ladder, however, form a bound
pair. Fusing all of these findings together,  and considering not only a single
domain wall but rather a larger lattice, 
we obtain stripe configurations suggested by DMRG calculations \cite{WS}, mean-field theory, and other treatments.
As has long been emphasized, DMRG computations are performed with open boundary conditions. These boundary conditions favor
(and, in some cases, may trigger) the formation of stripes and other inhomogeneities. Thus, stripes as seen by DMRG
may be argued to be stem from boundary effects. 
The DMRG results with open boundary conditions have been analyzed via bosonization and other means \cite{Ian}.
Notwithstanding the precise character of the pristine lowest energy states and the importance
of boundary effects in DMRG, similar patterns (including other geometries- those of diagonal and horizontal site-centered stripes) are found by numerous other means of analysis, e.g., \cite{corboz}. Our analysis illustrates that stripes may be stabilized by kinetic constrained. That is not to say
that these are the only possible low energy states. 

The arguments that we employed
in the current work were very pedestrian 
and general. Throughout we invoked 
the sublattice parity principle
and the related staggering
potentials. Our findings are further 
supported by numerical 
calculations on the $t$-$J$
model. We
give a very simple physical interpretation for
the stripes found by DMRG and other methods.

\begin{figure}
{\centering \resizebox*{0.7\columnwidth}{!}{\includegraphics{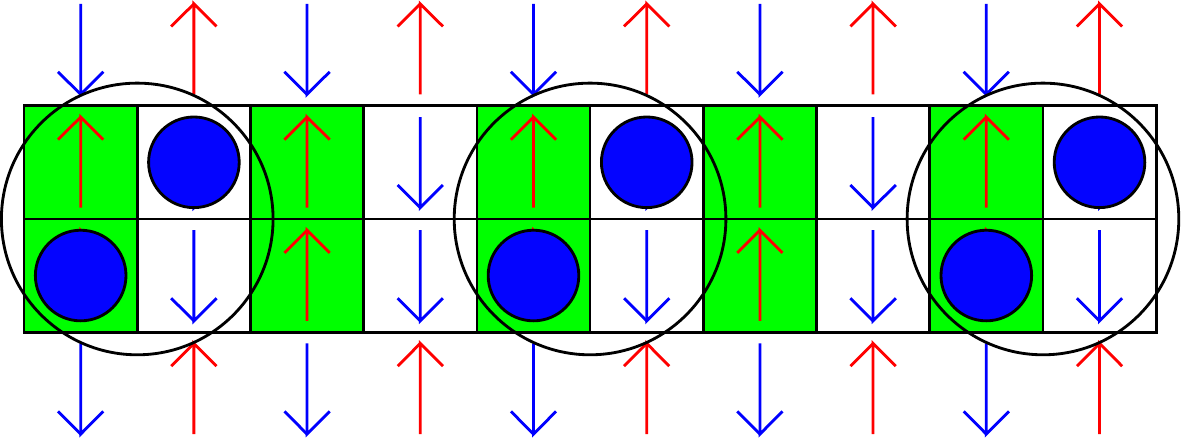}} \par}
\caption{For a quarter-filled stripe, the picture results.}
\end{figure}

\begin{figure}
{\centering \resizebox*{1\columnwidth}{!}{\includegraphics{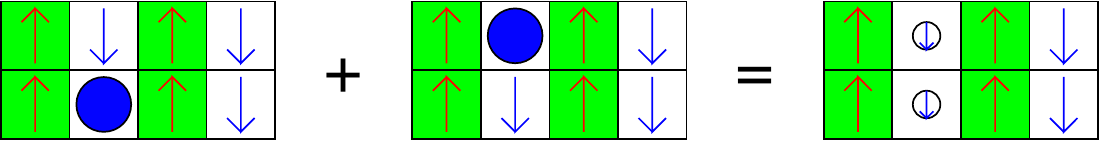}} \par}
\caption{The hole has equal amplitude to be on the upper or lower leg of the
ladder. Therefore, the groundstate is a linear superposition of a
hole and a spin down.}
\label{DmRg}
\end{figure}

A hole is on the stripe will delocalize by the hybridization
of the upper and lower leg. If we place a hole on the upper leg, 
the hole will almost immediately jump to the lower leg and 
vice verse. In numerical calculations, we will therefore 
find the linear superposition of a hole
and the electron depicted in 
Fig. (\ref{DmRg}).

\begin{figure}
{\centering \resizebox*{!}{5cm}{\includegraphics{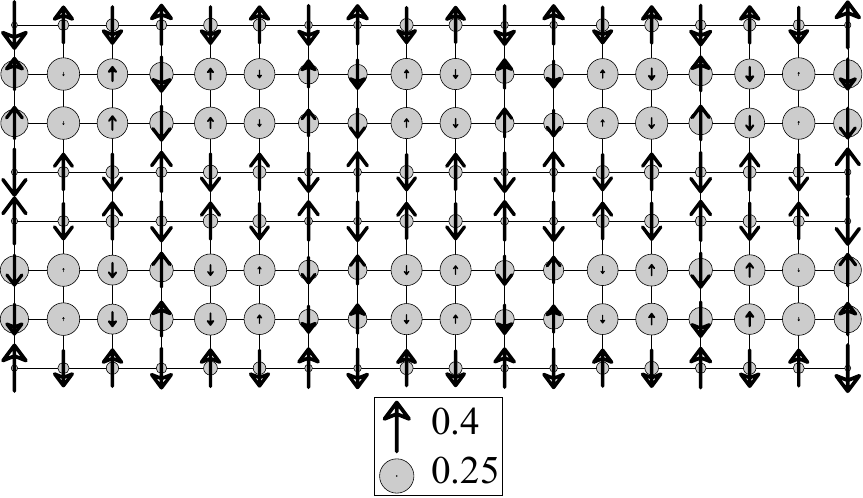}} \par}
\caption{The DMRG calculation of the $t$-$J$ model by White and
Scalapino \cite{WS}. This and similar calculations typically employ
open or cylindrical boundary conditions.}
\label{figure: White and Scalapino DMRG calculation result}
\end{figure}


White and Scalapino first employed the DMRG technique on the $t$-$J$
model \cite{WS}. 
Fig. (\ref{figure: White and Scalapino DMRG calculation result})
shows one of their well known results. We see that the
bond centered stripes are very narrow. They look like a ferromagnetic
seam in an antiferromagnet. The same result can be obtained in mean-field
theory \cite{Marcot}. 

The DMRG calculations of White and Scalapino and those of others since vividly 
suggested how spin and charge may nestle. Much emphasis was placed on the fact that the
holes are on next-nearest neighbor sites. On the opposite
diagonal, the spins (which are on the same sublattice) have a very
strong antiferromagnetic bonding. In effect, the bound pair is creating
a flip in the antiferromagnetic lattice. This might also explain why
in the exact calculations and the Monte-Carlo results the binding
energy would decrease as a function of the lattice size. Basically,
the hole-pair is creating an antiphase boundary, whose energy is increasing
as a function of lattice size. That this is indeed a good representation
of stripes follows from DMRG calculations by White and Scalapino
and other computations.

The reader is urged to focus on the central part of Fig. (\ref{figure: White and Scalapino DMRG calculation result}). (Here,
the spin and charge texture towards the boundaries are 
more contorted by the open boundary conditions employed.)
We claim that the reason that the mean-field \cite{Marcot} and the
DMRG calculation are so similar has to do with the fact 
that the antiferromagnet
is very strongly ordered. The stripes are so narrow because of {\it kinetic driven attraction} or {\em dynamical
confinement}. Although we focused on bond-centered stripes, similar considerations rationalize other stripe geometries.

\section{Longer Range Kinetic and Coulomb Terms}
\label{added}

Let us summarize the assumptions 
that we have invoked, so far, 
in our analysis:

(i) We assumed, self consistently, the existence 
of an antiphase domain wall having the geometry 
of a two leg ladder. Sublattice parity
($Z_{2}$) order was assumed to prevail
throughout the entire system.

(ii) Albeit the relatively minute
energy difference by which the lowest
lying pair states 
were found to be favored over single hole
states, we assumed that 
the stripe was composed entirely
of pairs.

It should be noted (especially in 
the context of assumption (ii)),
that the small energy differences
we found separating various 
contending states (as well
as those separating, say,
bond centered stripes from
site centered stripes) are
very susceptible to additional
terms in the Hamiltonian.

As we emphasized in the Introduction, the Hubbard model and its $t$-$J$ cousin
are only models. There are myriad very
important effects that it does not
include which could easily shift 
the balance between various 
nearly degenerate contending
states. 

Coulomb effects (which are much greater
importance here than elsewhere 
given the poor screening in
these materials) enhance and stabilize
stripe order: a uniform charge density
order is strictly forbidden by the
divergent Coulomb penalty that it will
incur. This point has been emphasized 
by Emery, Kivelson, and coworkers \cite{steve,low}.
Fourier transforming the Hamiltonian and looking
for the minimizing waveumbers,
we are able to see how stripe like charge 
density modulations will evolve \cite{me}
once lattice effects are taken 
into account.

Even if the stripe correlations found 
by DMRG and other calculations are 
triggered by the application of open 
boundary condition effects,
when long range Coulomb interactions
are introduced, charge stripe order
will be further stabilized.
Given the natural coupling between
spin and charge \cite{oron}, this will
further enhance the sublattice
parity flips across the 
stripe that we assumed
from the outset (assumption
(i)). 
 
Less emphasized are the role of
higher order kinetic terms. 
These can easily tip the balance:
a next nearest neighbor hopping
increases pairing significantly.
Within the pure $t$-$J$ model (with a 
vanishing direct diagonal hopping amplitude $t^{\prime}=0$) pairing 
correlations are infinitesimal and are further
frustrated by the $\pi$ phase shift across the domain
wall. Numerically, $t^{\prime}$ which 
allows holes to move on the same sublattice 
enhances pairing dramatically. Furthermore, 
numerically, a negative $t^{\prime}$ is seen to 
favor stripe formation \cite{Ogata}.

\section{Conclusion}

The principle objective of the current work was to demonstrate that stripe ordering
and pairing tendencies in repulsive doped Hubbard type models may result from kinetic considerations.
Our analysis indeed illustrated that {\it stripe order 
self-consistently emerges from the existence of a strong antiferromagnetic background
that forces holes to move on the same sublattice}. 
We demonstrated that holes move to antiferromagnetic domain
walls and effectively form two-leg ladders. This \emph{dynamical
confinement} of holes onto the stripe is caused by the sublattice
structure of the antiferromagnet and the increase in kinetic energy
on a domain wall. The effective two-leg ladders still feel the influence
of the surrounding antiferromagnet in terms of a staggered boundary
potential. As shown by Krotov, Lee, and Balatsky\cite{KLBafboundary},
this increases the tendency to superconductivity. The reason for this
is that holes on the stripe form real-space pairs. 
In principle, our ideas might be tested 
for other bipartite lattices in which 
a background N\'eel order may 
favor hopping between sites
on the same sublattice.

In a related work \cite{us2}, we illustrated that these real-space
pairs can be mapped to an effective one-dimensional 
XXZ model in a transverse field. This enables us to discuss
the filling fractions of stripes.

\begin{acknowledgments}
The authors would like to thank Wim van Saarloos for
encouragement and discussions and Jan Zaanen for critical reading of
the manuscript and for subsequently
pointing out to us \cite{pryadko,strings}. We were 
partially supported by the Foundation of Fundamental 
Research on Matter (FOM), which is sponsored 
by the Netherlands Organization of 
Pure research (NWO).  ZN gratefully
acknowledges partial support by the NSF under
grant no. DMR 1411229. 

\end{acknowledgments}

\appendix

\section{The sublattice parity principle}
\label{sub_prin}

The sublattice parity principle amounts to the  
assumption that a hole in an antiferromagnet is free and can move to all
of its {\em next} nearest but not to its direct
neighbors.
This principle is based on and follows from the  
sublattice structure of antiferromagnetic 
order.
We will provide phenomenological proof for this assumption
by examining the dispersion relations obtained by
very detailed numerical works.

\subsection{Spin flips}
It is well known that the Hubbard model for
a half-filled system in the large \( U \) limit 
leads to an effective Heisenberg model,

\begin{equation}
H_{\mbox {Heisenberg}}=\sum _{\langle ij\rangle
}\frac{J_{\perp}}{2}(S_{i}^{+}S^{-}_{j}+S^{-}_{i}S_{j}^{+})+J_{z}S^{z}_{i}S^{z}_{j},
\end{equation}
with $J_{\perp} = J_{z}=J$.

The last (Ising) term of the Heisenberg Hamiltonian
wants to make the spins on neighboring lattice sites point in 
opposite directions. If only this term is present,
the groundstate will be a perfect Ising antiferromagnet with
long-range (N{\'e}el)
order: the lattice can be subdivided in an up sublattice and a down
sublattice. However, the first (XY) term in the Hamiltonian can undo 
this order by flipping two neighboring, opposite spins.
In principle this term can completely destroy the long range order
and indeed it does so in one dimension. In dimensions $d\ge 3$ 
clear sublattice order prevails. 
A large amount of effort
in numerical calculations and the very important analysis of the
non-linear sigma model by Chakravarty, Halperin and Nelson\cite{CHN}
have shown that the ground-state of the two-dimensional Heisenberg
model (and thus of the undoped Hubbard model) is a long range antiferromagnet
for which there exists a two-sublattice structure. The groundstate
of the antiferromagnet is not exactly given by the classical 
N{\'e}el state which one would expect from the Ising model. The spin-flip term
leads to a finite density of flipped spins (approximately 20\%) which
give rise to a lowering of the magnetization. It also leads to spin-waves.
These spin-waves destroy the long-range antiferromagnetic order at
any finite temperature. However, one can define a correlation length
and on distances smaller than this length, we can still speak about
local antiferromagnetic order. For \( T>0 \), we have local order
with a correlation length that decreases with increasing temperature.
Therefore, thinking in terms of an Ising model is not completely incorrect
as long as one keeps in mind that there is a finite density of flipped
spins in the lattice. In the remainder of this article we will neglect
these spin flips.

\begin{figure}
{\centering \resizebox*{1\columnwidth}{!}{\includegraphics{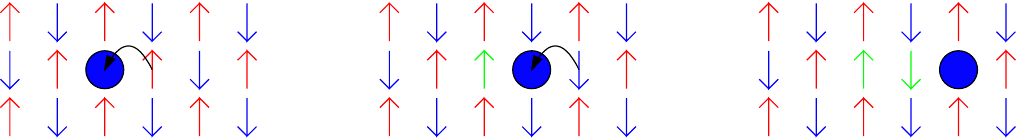}} \par}
\caption{If a single hole in an antiferromagnet moves by nearest neighbor
hopping of surrounding spins, it cannot move without creating a string
of flipped spins. For small values of \protect\( J_{\perp}\protect \),
there are no spin flips that can destroy this string. This leads to
a linearly increasing confining potential and the eigenfunctions are
Airy wave functions. Because of this string, a single hole cannot
easily propagate through an antiferromagnet by nearest neighbor hops.}
\label{figure: stringstates in an AF.}
\end{figure}

A moment's reflection reveals that $[H, S_{z}^{tot}]=0$
for the Hubbard model (a property inherited
to its descendants), and net magnetization 
is strictly preserved.
Consequently, if N{\'e}el order prevails also
when a single hole is introduced, hole motion
is rigorosly restricted to one sublattice: if a
hole could indeed hop to its neighboring site
on the opposite sublattice, the magnetic quantum
number $(S_{z}/\hbar)$ would remain unaltered. On
the other hand the two low energy ``vacuum'' states
corresponding to the injection
of a hole on the two different sublattices
have magnetic quantum numbers $(S_{z}/\hbar)$
differing by $\pm 1$. These states obviously
cannot be connected by any of the 
magnetization conserving processes of $H$.
Myriad analytical treatments of the $t$-$J$ and
Hubbard models were  aware of this selection
rule and have computed transition 
matrix elements for a single hole between sites of the 
same sublattice. For one example amongst 
many see \cite{highd}.

A well known argument leads to the conclusion that a single hole cannot
propagate freely in an antiferromagnet. As shown in Fig.
(\ref{figure: stringstates in an AF.}), whenever a non backtracking
single hole moves from one sublattice to
the other, it creates a string of flipped spins in its wake. 
This leads to a linear confining potential
that strongly inhibits hole motion between different sublattices.

\begin{figure}
{\centering \resizebox*{1\columnwidth}{!}{\includegraphics{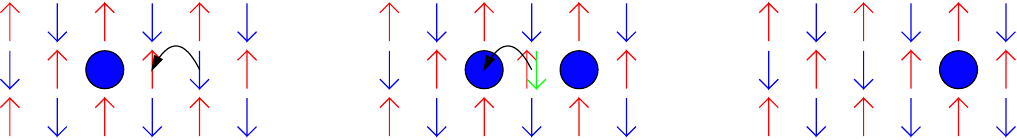}} \par}
\caption{A single hole can still move through an antiferromagnet without generating
string states by hopping in steps of two. This way it stays on the
same sublattice. For this it has to go through an intermediate state
with a double occupied site. This intermediate state costs an energy
\protect\( U\protect \). This leads to movement of the hole on the
same sublattice with an amplitude \protect\( \propto t^{2}/U\protect \).}
\label{figure: Hole moving in steps of 2 through an AF}
\end{figure}

This, however, is not the only process that can take place. Fig.
(\ref{figure: Hole moving in steps of 2 through an AF}) shows that
it is also possible for a hole to move two steps 
without creating a confining string potential. This is a second order process
in perturbation theory. First we create a second hole two sites away
from the first hole. We do this by moving that electron one site closer,
next to the first hole. This gives rise to a double occupied intermediate
state with a high energy \( U \). Then, we let the same electron
move again, now removing the first hole. The
amplitude for the total process is given by \( \sigma =\frac{t^{2}}{U} \),
much smaller than \( t \), the first order hopping
amplitude. However, the final state has exactly the same energy as the
starting state. There are eight sites to which a single hole can hop in
this fashion. 
Note that throughout the entire process, the 
magnetic quantum number in Fig. (\ref{figure: Hole moving in steps of 2
through an AF}) is preserved, e.g. $S_{z}^{tot} = \hbar/2$ if
perfect N{\'e}el order prevails everywhere around the 
fragment displayed in the figure for a lattice with an
even number of sites. Within the t-J and $t-J_{z}$ approximations
to the Hubbard model, same sublattice 
hops are more masked. In
the t-J model, only a third order process
links a hole to one of its next nearest neighbors
on the same sublattice. Albeit holding
for models derived from
the Hubbard model on bipartite lattices,
in many instances the sublattice parity 
principle becomes much more alive and 
transparent within the original Hubbard 
model itself.

\subsection{Numerical dispersions}
\begin{figure}
{\centering \resizebox*{1\columnwidth}{!}{\includegraphics{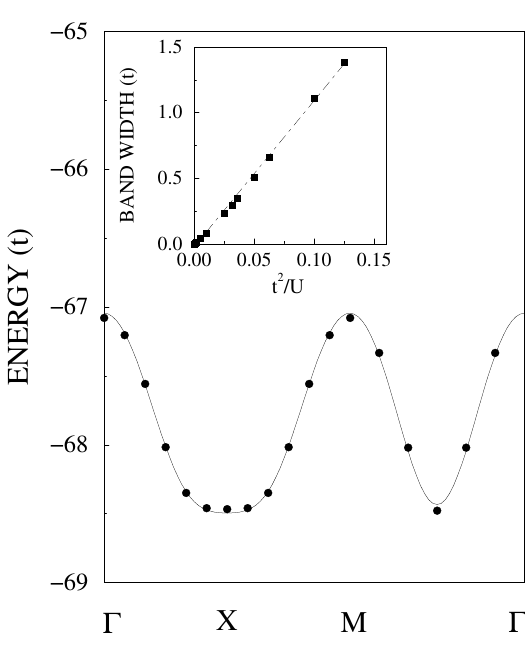}} \par}
\caption{Quasiparticle band structure for a single hole on \protect\( 12\times 12\protect \)
clusters of the square lattice with periodic boundary conditions and
\protect\( U=8t\protect \). The solid line corresponds to the fitted
dispersion relation (see text). The inset shows the bandwidth as a
function of \protect\( t^{2}/U\protect \) for \protect\( U\ge 8t\protect \);
the fitted straight line is \protect\( -0.022t+11.11t^{2}/U\protect \).
From Louis, Guinea, L{\'o}pez Sancho and Verg{\'e}s \cite{Louis}.}
\label{figure: dispersion relation for single hole + bandwith versus J}
\end{figure}

Simulations of the Hubbard and Heisenberg model with a single
hole have found a discrete number of sharp peaks in the spectrum which
could be identified with string-states. The longer the string, the
higher the energy of the peak. Because there is a gap between the
first and second peak, we can confine our attention to the lowest
peak. This peak has a finite quasi-particle weight. The dispersion
of this peak as a function of \( k \), tells us about the effective
movement of a hole through an antiferromagnet. A numerical
result from Louis {\em et. al.} \cite{Louis} 
for this dispersion relation in the Hubbard model is shown
in Fig. (\ref{figure: dispersion relation for single hole + bandwith versus J}).

The central result of numerous investigations is that they all found
that a sharp quasiparticle peak appears at the bottom of a broad continuum
of the hole spectrum. These quasiparticle poles form a coherent hole
band with a width of order of $2J$ over a wide range of $J/t$, and the
coherent propagation is made possible by ``healing'' a string
of flipped spins by quantum fluctuations.

One can easily Fourier 
transform the numerical low energy hole dispersion
relation $\epsilon(\vec{k})$ to unveil the important
real space quasiparticle motions and their 
respective amplitudes. If we make the simplifying
assumption that the hole is the quasiparticle
then we find that essentially 
all of the low energy weight is distributed 
amongst the next nearest neighbor motions linking
the hole to its sublattice. This is a consistent logical
outcome of a strong local N{\'e}el order fused with the
fact that the magnetic moment is a conserved quantum number.
That the nearest states on the same sublattice (and
those further away)
have the highest weight could hardly be 
surprising. Hopping amplitudes to sites which
are further and further away 
on the same sublattice drops significantly
with distance \cite{Louis}.

Stated alternatively, if on the two dimensional 
lattice, the hopping amplitude to any of the 
four colinear sites (twice (up, down, right or left)) 
is \( t_{0} \) and if motion to any of the four
diagonal sites has amplitude \( t_{2} \))
then the dispersion relation will read

\begin{equation}
\label{equation: dispersion for hole in AF}
\epsilon _{k}=\epsilon _{0}-4t_{2}\cos k_{x}\cos k_{y}-2t_{0}(\cos 2k_{x}+\cos 2k_{y}).
\end{equation}

\begin{figure}
{\centering \resizebox*{1\columnwidth}{!}{\includegraphics{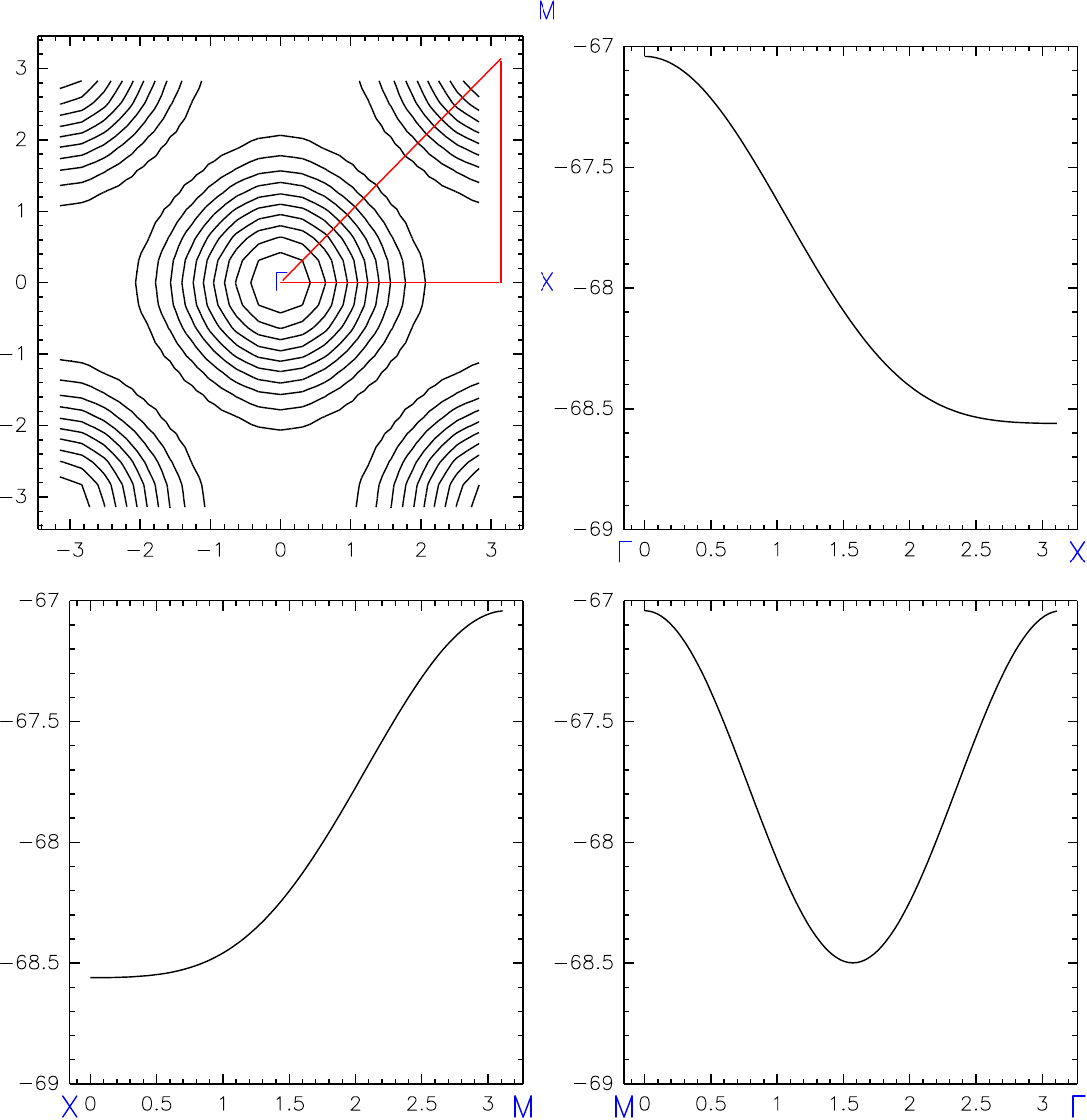}} \par}
\caption{Dispersion relation for a single hole in an antiferromagnet according
to the simple view that the hole is a quasi-particle that moves on
one sublattice by hopping in steps of two (equation \ref{equation: dispersion for hole in AF}).
Here we employ \protect\( \epsilon _{0}=-68.15t\protect \), \protect\(
t_{2}=-1.52 \frac{t^{2}}{U}\protect \) and \protect\( t_{0}= - 0.70\frac{t^{2}}{U}\protect \) with \protect\( U=8t\protect \).
\label{figure: theoretical dispersion single hole}}
\end{figure}

From lowest order perturbation theory \cite{highd}, we immediately expect
\( t_{2}=2 t_{0}= - {\cal{O}}( \frac{t^{2}}{U}) \) where the 
relative factor of two originates from the two paths by which 
we may reach diagonal sites by two consecutive hops
as comparde to the single two step
route to longintudinal next nearest
neighbors.
As illustrated in Fig. (\ref{figure: theoretical dispersion single hole}), 
renormalized hopping amplitudes of the same order of magnitude 
${\cal{O}}(\frac{t^{2}}{U})$
reproduce the detailed and tedious numerical fits of
Fig. (\ref{figure: dispersion relation for single hole + bandwith versus
J}) remarkably well. More generally, if one
Fourier transforms the elaborate data 
encoded in Fig. (\ref{figure: dispersion relation for single hole + bandwith versus J})
we find that the bulk of the Fourier weight corresponds
to next to nearest neighbor motions with hopping
amplitudes of the same order as anticipated from
lowest order perturbation theory \cite{Louis}.

In accordance with expectations
from perturbation theory, it is indeed found numerically
that diagonal ({\em nodal}) hopping is almost twice 
as large in amplitude as compared to
longitudinal hopping (a property which might easily
transcend to larger length scales by virtue of
many of the small length scales 
in these materials) with both amplitudes 
of order ${\cal{O}}(\frac{t^{2}}{U})$.
In Fig. (\ref{figure: theoretical dispersion single hole}),
we show the theoretical dispersion curves coming from equation (\ref{equation: dispersion for hole in AF})
for \( t_{0}= -0.70 \frac{t^{2}}{U} \) and \( t_{2}= -1.52 \frac{t^{2}}{U} \)
and \( \epsilon _{0}=-68.15t \). These curves capture all the essential
features of the numerical dispersion curve shown in Fig. (\ref{figure: dispersion relation for single hole + bandwith versus J}).

It is obvious that the Brillouin zone has doubled as a result of the
fact that the hole is moving in steps of two through the lattice.

The very good fit obtained to the dispersion
curve of a single hole in an antiferromagnet by introducing
the three parameters
(\( \epsilon _{0} \), \( t_{0} \) and \( t_{2} \)) suggests that the
complicated problem of a single hole in an antiferromagnet can be
reduced to a single particle problem. This
is a major simplification. If we really can neglect
spin-flips and string-states, the single hole problem has been reduced
to a single-particle problem. The spins have become static and their
only function is to create a checkerboard like sublattice structure
that keeps the hole moving on only one sublattice by forcing it to
move in steps of two. Large \( J \) means in principle both large
\( J_{z} \) and \( J_{\perp} \). A large \( J_{\perp} \) leads to spin-flips
being relatively important and this will restore the string-states.
Increasing the value of \( J \), and thus of \( J_{\perp} \), leads
to spin-flips which will restore a string to the normal antiferromagnet.
However, the relevant region of the Hubbard model and the $t$-$J$
model is the region of small values of \( J \).

Stated more generally, the constraint on hole motion (equivalent to
the assumption of N{\'e}el order) may be 
regarded as a low energy selection rule 
on the matrix elements of the
effective Hamiltonian. The total $z$ component
of spin (sublattice parity order) is a conserved 
quantum number and transitions between different
sublattice states are banned. Whenever 
N{\'e}el order is strong, the states
are effectively separated into two 
disjoint Hilbert spaces each for a 
different sublattice parity 
(magnetic $S_{z}$) quantum number. 
As we demonstrate in the main
text, stripe order may be interpreted 
as an immediate consequence of sublattice parity 
quantum numbers.

\end{document}